\documentclass[10pt]{article}
\usepackage[hyphens]{url}
\usepackage{hyperref}

\usepackage{graphicx}
\usepackage[ruled,linesnumbered]{algorithm2e}

\usepackage{booktabs} % For formal tables
\usepackage{authblk}
\usepackage{tabu}
\usepackage{adjustbox}
\usepackage{color, colortbl}
\usepackage{subcaption} 
\usepackage{amsthm}
\usepackage{amsmath}
\newtheorem{quest}{Question}
\usepackage{pdflscape}
\usepackage{enumitem}

\definecolor{Gray}{gray}{0.9}
\definecolor{White}{rgb}{1,1,1}
\definecolor{LightCyan}{rgb}{0.88,1,1}

\usepackage[space]{grffile}

\makeatletter
\g@addto@macro{\thm@space@setup}{\thm@headpunct{:}}
\makeatother% Copyright
\begin{document}
\title{Using Data Science to Understand the Film Industry's Gender Gap}
% \titlenote{Produces the permission block, and
% \subtitle{Extended Abstract}
% \subtitlenote{The full version of the author's guide is available as
%   \texttt{acmart.pdf} document}

% \author{Dima Kagan}
% \affiliation{%
%   \institution{Ben-Gurion University of the Negev}
% }
% \email{kagandi@post.bgu.ac.il}

% \author{Thomas Chesney}
% \affiliation{%
%   \institution{Nottingham University Business School}
% }
% \email{thomas.chesney@nottingham.ac.uk}

% \author{Sean Mcnaughton}
% \affiliation{%
%   \institution{to do}
% }
% \email{todo@email.com}

% \author{Michael Fire}
% \affiliation{%
%   \institution{Ben-Gurion University of the Negev}
% }
% \email{mickyfi@post.bgu.ac.il}

\author[1]{Dima Kagan\thanks{kagandi@post.bgu.ac.il}}
\author[2]{Thomas Chesney\thanks{thomas.chesney@nottingham.ac.uk}}
\author[1]{Michael Fire\thanks{mickyfi@post.bgu.ac.il}}
\affil[1]{Department of Software and Information Systems Engineering, Ben-Gurion University of the Negev}
\affil[2]{Nottingham University Business School}
% \affil[3]{The eScience Institute, University of Washington}
% \renewcommand{\shortauthors}{B. Trovato et al.}

    \maketitle

\begin{abstract}
Data science can offer answers to a wide range of social
science questions. Here we turn attention to the portrayal of women in
movies, an industry that has a significant influence on society, impacting
such aspects of life as self-esteem and career choice. To this end, we
fused data from the online movie database IMDb with a dataset
of movie dialogue subtitles to create the largest
available corpus of movie social networks (15,540 networks). Analyzing
this data, we investigated gender bias in on-screen female characters over
the past century.

We find a trend of improvement in all aspects of women`s roles in movies, including a constant
rise in the centrality of female characters. There has also been an increase in the
number of movies that pass the well-known Bechdel test, a popular---albeit flawed---measure of women in fiction. Here we propose a new and
better alternative to this test for evaluating female roles in movies. Our study introduces fresh data, an open-code framework, and novel techniques that present new opportunities in the research and analysis of movies.

\end{abstract}


	\providecommand{\keywords}[1]{\textbf{Keywords:} #1}


\section{Introduction}
\label{sec:int}

\begin{landscape}
\begin{figure}
\refstepcounter{figure}\label{fig:infog}

\centering
\includegraphics[width=1\linewidth]{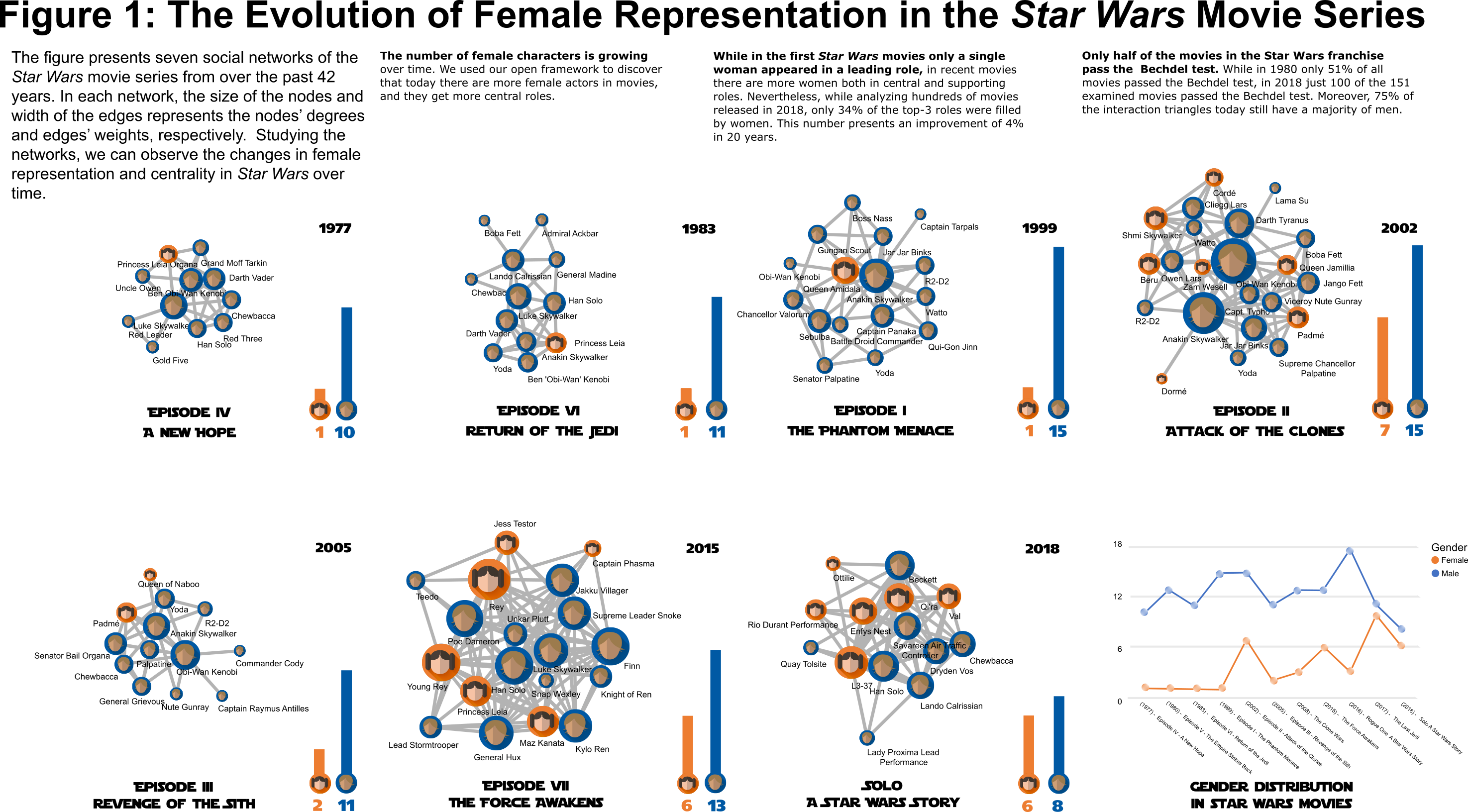}
\end{figure}

\end{landscape}
The film industry is one of the strongest branches of the media, reaching billions of viewers worldwide~\cite{PowerPoi43:online,UNICAR2049:online}.
Now more than ever, the media has a major influence on our daily lives~\cite{silverstone2003television},
significantly influencing how we think \cite{entman1989media}, what we wear \cite{wilson1998self}, and our self-image \cite{polce2001adolescent}. In particular, the representation of women in media has an enormous influence on society.
As just one example, a new study shows that ``women who regularly watch \textit{The X-Files} are more likely to express interest in STEM, major in a STEM
field in college, and work in a STEM profession than other women in the sample''~\cite{ScullyEf96:online}.

Movies are the fulfillment of the vision of the movie director, who controls all aspects of the filming.
It is well known that movie directors are primarily white and male~\cite{Inclusio46:online}. With such a gender bias, it is not surprising that there is a male gender dominance in movies~\cite{fullstud89:online}.
Studies from the past two decades have confirmed that women in the film industry are both underrepresented~\cite{Womenrem83:online, lauzen2018boxed} and portrayed stereotypically~\cite{wood1994gendered}.
A recent study found that the underrepresentation is so sizeable that there are twice as many male speaking characters as female in the average movie~\cite{2017Itsa47:online}.

While the gender gap in the film industry is a well-known issue \cite{ 2017Itsa47:online,Onefemal15:online,WomenAnd61:online, lauzen2018boxed,wood1994gendered}, there is still much value in researching this topic.
Most previous gender studies can be categorized into two types:
the first type offers simple statistics from the data to emphasize the gender gap~\cite{lauzen2018boxed}; and the second type introduces more advanced analytical methods, yet generally uses only a small amount of data~\cite{agarwal2015key,garcia2014gender}.

In this study, 
we present \textit{Subs2Network}, a novel algorithm to construct a movie character's social network.
We demonstrate possible utilizations of \textit{Subs2Network} by employing the latest data science tools to comprehensively analyze gender in movies (see Figure \ref{fig:infog}\footnote{The \href{https://www.behance.net/gallery/17998561/Star-Wars-Long-Shadow-Flat-Design-Icons}{Star Wars icons} were created by \href{https://www.behance.net/creativeflip}{Filipe de Carvalho} and are licensed under CC BY-NC 4.0}). This is the largest study to date that uses social network analysis (SNA) to investigate the gender gap problem in the film industry and how it evolved.

The study's primary goals are to answer the following four questions:
\begin{quest}
Are there movie genres that do not exhibit a gender gap?
\end{quest}
\begin{quest}
What do characters' relationships reveal about gender, and how has this changed over time?
\end{quest}
\begin{quest}
Are women receiving more central movie roles today than in the past?
\end{quest}
\begin{quest}
How has the fairness of female representation in movies changed over the years?
\end{quest}

To answer these questions, we first analyzed movie subtitles using text-processing algorithms and a list of movie characters' names (see Figure~\ref{fig:flow}). We then developed \textit{Subs2Network} to construct a movie character's social network. We created an open-source code framework to collect and analyze movie data, and we used this framework to construct the largest existing open movie social network dataset that exists today. 

Using the constructed movie social networks, we extracted dozens of topological features that characterized each movie.
By analyzing these features, we could observe the gender gap across movie genres and over the last 99 years.
Moreover, by utilizing the dataset, we developed a machine-learning classifier, which is able to assess, with \textit{precision at 200} of 0.94, how fairly women are represented in movies (i.e., if a movie passes the Bechdel test \cite{bechdel1985rule}).

Our results demonstrate that
%it is possible to classify a movie character's gender solely by the character's metadata. In addition, 
in most movie genres there is a statistically significant difference between men and women in centrality features like \textit{betweenness} and \textit{closeness}.
These differences indicate that men are getting more central roles in movies than women (see Figures \ref{tab:deg} and \ref{tab:page}, and Section \ref{sec:results}).
Another sign of the underrepresentation of women in movies is found by analyzing interactions among three characters: only 3.57\% of the interactions are among three women, while 40.74\% are among three men.
These results strengthen previous studies` results that women play fewer central roles~\cite{agarwal2015key,lauzen2018boxed}, and indicates that on average women have more minor roles.
Our results highlight how and where gender bias manifests in the film industry and provides an automatic way to evaluate it over time.

The key contributions presented in this paper are fivefold:
\begin{itemize}
    \item A novel algorithm (see Section~\ref{sec:method}) which utilizes movie subtitles and character lists to automatically construct a movie's social network (see Section~\ref{subsec:snet} and Figure~\ref{fig:flow}).
    
    \item The largest open movie social network dataset, 21 times larger than the previous dataset~\cite{kaminski2018moviegalaxies} (see Section~\ref{subsec:data}).
    Our dataset contains 15,540 dynamic networks of movies (937 of these networks are networks  of biographic movies, which have information about real world events). 
    \item An open-source framework for movie analysis.
    The code contains a framework to generate additional social networks of movies,  facilitating research by creating and analyzing larger amounts of data than ever before.
    
    \item A machine-learning classifier that can predict if a movie passes the Bechdel test (see Section~\ref{subsec:bechdel}) and can evaluate the change in gender bias in thousands of movies over several decades (see Section~\ref{sec:results}). 
    
    \item Our new and alternative automated Bechdel test to measure female representation in movies. This new test overcomes the weaknesses of the original Bechdel test.
\end{itemize}

% \begin{figure}[ht!]
%   \centering
%     \includegraphics[width=1\linewidth]{The Social.png}

%   \caption{Social network of the movie \textit{The Social Network}.}
%   \label{fig:socialnet}

% \end{figure}
Our study demonstrates that inequality is still widespread in the film industry. In movies of 2018, a median of 30\% women and a mean of 33\% were found in each movie's top-10 most central roles. That being said, there is evidence that the gender gap is improving (see Figure \ref{fig:trend}).
%grows every year.
%If this trend will continue, the gap between genders will be a story of the past.

The remainder of this paper is organized as follows:
In Section \ref{sec:rw}, we present an overview of relevant studies.
In Section \ref{sec:method}, we describe the datasets, methods, algorithms, and
experiments used throughout this study. 
In Section \ref{sec:results}, we present our results.
Then, in Section \ref{sec:dis}, we discuss
the obtained results.
Lastly, in Section \ref{sec:con}, we present our conclusions from this study and offer future research
directions.

\section{Related Work}
\label{sec:rw}

\subsection{Movie Social Networks}
In the past decade, the study of social networks has gained massive popularity.
Researchers have discovered that social network analysis techniques can be used in many domains that do not have explicit data with a network structure.
One such domain is the film industry. Researchers have applied social network analysis to analyze movies, gaining not only
 new insights about specific movies but also about the film industry in general.
For example, using social networks makes it possible to empirically analyze social ties between movie characters. 

In 2009, Weng et al.~\cite{weng2009rolenet} presented RoleNet, a method to convert a movie into a social network.
The RoleNet algorithm builds a network by connecting links between characters that appear in the same scene.
RoleNet is based on using image processing for scene detection and face recognition to find character appearances.
Weng et al. evaluated their method on 10 movies and 3 TV shows. 
The method was used to perform semantic analysis of movies, find communities, detect leading roles, and determinine story segmentation. 

In 2012, Park et al.~\cite{park2012social} developed Character-net, another method to convert movies to networks.
Character-net builds the social network based on dialog between characters, using script-subtitle alignment to extract who speaks to whom in the scene.
Park et al. evaluated their method on 13 movies~\cite{park2012social}.
Similar to RoleNet, Character-net was used to detect leading roles and to cluster communities.

In 2014, Agrawal et al.~\cite{agarwal2014parsing} presented a method for parsing screenplays by utilizing machine-learning algorithms instead of using regular expressions.
Their study showed that the parsed screenplay can be used to create a social network of character interactions.
In 2015, Tran and Jung~\cite{tran2015cocharnet} developed the CoCharNet, a method which adds weight to a link in the interaction network, where the weight is a function of the number of times two characters appear together.
Tran and Jung used CoCharNet to evaluate the importance of characters in movies.
They demonstrated that network centrality features such as closeness centrality, betweenness centrality, and weighted degree can be used to classify minor and main characters in a movie.
For instance, they detected the main characters using closeness centrality with a precision of 74.16\%.

In 2018, Lv et al.~\cite{lv2018storyrolenet} developed an algorithm to improve the accuracy of creating social networks of movies.
They presented StoryRoleNet, which combines video and subtitle analysis to build a more accurate movie social network.
The subtitles were used to add additional links that the video analysis might miss.
Similar to RoleNet and Character-net, Lv et al. used the movie social networks to cluster communities and to detect important roles.
They evaluated the StoryRoleNet method on 3 movies and one TV series, for which they manually created  baseline networks~\cite{lv2018storyrolenet}.

Also in 2018, a dataset from Moviegalaxies~\cite{kaminski2018moviegalaxies}\footnote{\url{http://www.moviegalaxies.com}} was released.
Moviegalaxies is a website that displays social networks of movie characters.
The dataset contains 773 movie social networks that were constructed based on movie scripts.
However, Moviegalaxies did not disclose the exact methods which were used for the construction of the networks.

\subsection{Evaluating the Gender Gap}
In recent years, there have been many studies that attempt to evaluate the gender gap between males and females across various domains \cite{jia2016women, lariviere2013bibliometrics,lauzen2018boxed, wagner2015s}.
For example, in 2018 the World Bank evaluated that the costs of gender bias are vast; gender inequality results in an estimated \$160.2 trillion~loss in human capital wealth \cite{Unrealiz38:online}.

Over the years, researchers have discovered many manifestations of the gender gap in our society.
Lariviere et al.~\cite{lariviere2013bibliometrics} discovered that scientific articles with women in dominant author positions receive fewer citations.
Wagner et al.~\cite{wagner2015s} observed that men and women are covered equally on Wikipedia, but they also discovered that women on Wikipedia are portrayed differently from men.
Jia et al.~\cite{jia2016women} found that in online newspapers, women are underrepresented both in text and images.  

The state of women in the film industry is similar to other domains: women are underrepresented and badly portrayed~\cite{lauzen2018boxed, wood1994gendered}.
The \textit{Boxed In 2017-18} report~\cite{lauzen2018boxed} observed a 2\% decline in female major characters across all platforms, compared to the previous year.

To tackle the underrepresentation of women in movies in 1985, the cartoonist Alison Bechdel published a test in her comic strip \textit{Dykes to Watch Out For} to assess how fairly women are presented in filmed media.
The Bechdel-Wallace test \cite{bechdel1985rule} (denoted as the \textit{Bechdel test}) has three rules that a movie has to pass to be considered ``women friendly'':
\begin{enumerate}
    \item It has to have at least two women in it.
    \item The women have to talk to each other.
    \item The women must talk about something besides a man.
\end{enumerate}
Only 57\% of current movies pass this test.
To Bechdel's surprise, the media adopted her joke, and today it is a standard for female representation in movies~\cite{TheBechd75:online, ComicCon18:online, TheDolla99:online, mediares11:online,Oscars2038:online}.

The Bechdel test is also used by researchers.
In recent years, studies have utilized the test to evaluate gender bias in movies.
In 2014, Garcia et al.~\cite{garcia2014gender} quantified the Bechdel test and also applied it to social media. They joined YouTube trailers, movie scripts, and Twitter data, which resulted in 704 trailers for 493 movies and 2,970 Twitter shares. Garcia et al. created a social network of dialogues for these movies.
Additionally, they constructed a network of dialogues between Twitter users who discussed the trailers.
They mapped dialogues between men who were referring to women and between women who were referring to men.
This mapping was used to calculate the Bechdel score.
They found that trailers of movies which are male biased are more popular.
Also, they discovered that Twitter dialogues have a similar bias to movie dialogues~\cite{garcia2014gender}.

In 2015, Agarwal et al.~\cite{agarwal2015key} studied the differences between movies that pass and fail the Bechdel test.
Similar to Garcia et al., Agarwal et al. also constructed social networks using screenplays.
They created a classifier to automate the Bechdel test, which was trained on 367 movies and evaluated on 90.
In the evaluation, they discovered that network-based features perform better than linguistic features.
Additionally, they discovered that movies that fail the Bechdel test tend to have women in less central roles~\cite{agarwal2015key}.
With this being said, the Bechdel test has several major flaws.
The test does not take into account if women are represented stereotypically \cite{WhytheBe57:online}.
Additionally, there are movies that are considered feminist but do not pass the test \cite{22Movies88:online}.
Moreover, the test is considered to be a low threshold since a film can pass the test with a single line of dialogue between two women~\cite{mediares11:online}.

\subsection{Graph Features and Named Entity Recognition}
Data science tools and techniques have evolved rapidly in the past couple of years~\cite{donoho201550}. In this study, we primarily utilized data science algorithms from the domains of Natural Language Processing (NLP) and Social Network Analysis to computationally analyze movie content, movie social network structure, and how movie features change over time.

Namely, we used NLP to extract character names from the movie subtitles by utilizing Named Entity Extraction (NER) algorithms~\cite{nadeau2007survey}. We used both Stanford Named Entity Recognizer~\cite{finkel2005incorporating} and spaCy Python Package~\cite{spacy2} to find where characters appear in the text.

To match characters' names in the subtitles with characters' full names, we utilized FuzzyWuzzy~\cite{seatgeek4:online}, a Python package for fuzzy string matching.
Specifically, we used FuzzyWuzzy's \textit{WRatio}~\cite{fuzzywuz32:online}, a method for measuring the similarity between strings.
\textit{WRatio} uses several different preprocessing methods that rebuild the strings and compare them using Levenshtein distance~\cite{levenshtein1966binary}.
Also, \textit{WRatio} takes into account the ratio between the string lengths.
% Afterwards, the differences and intersection between the sets are calculated and the sets sorted alphanumerically.
% Later, Levenshtein distance calculated for the combinations of the sorted sets and the max value is returned.

After extracting the movie characters, we constructed the movie social networks and used various graph centrality algorithms, such as closeness, betweenness, degree centrality, and PageRank~\cite{brandes2005network} to identify the most central characters in each constructed movie network.

\section{Methods and Experiments}
\label{sec:method}
\subsection{Constructing Movie Social Networks}
\label{subsec:snet}

% \subsubsection{Algorithm}
One of this study's primary goals was to develop a straightforward algorithm that would construct the social network of character interaction within a given movie. We achieved this goal by utilizing movie subtitles\footnote{Many of the used movies' subtitles were created by crowd-sourcing, i.e., by people who volunteered to create the subtitle.} and a list of movie character names.
Namely, given a movie, we constructed the movie social network $G := <V,E>$, where $V$ is the network's vertices set, and $E$ is the set of links among the network's vertices. Each vertex $v \in V$ is defined to be a character in the movie. Each link $e:=(u,v,w) \in E$ is defined as the interaction between two movie characters $u$ and $v$, $w$ times.

\begin{figure}[ht!]
  \centering
  \begin{subfigure}[c]{0.29\linewidth}
    \includegraphics[height=3cm]{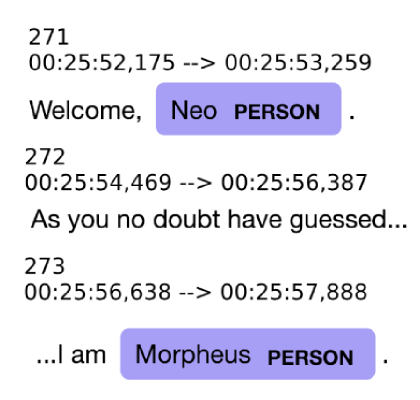}
    \caption{NER}
  \end{subfigure}
  \begin{subfigure}[c]{0.29\linewidth}
    \includegraphics[height=3cm]{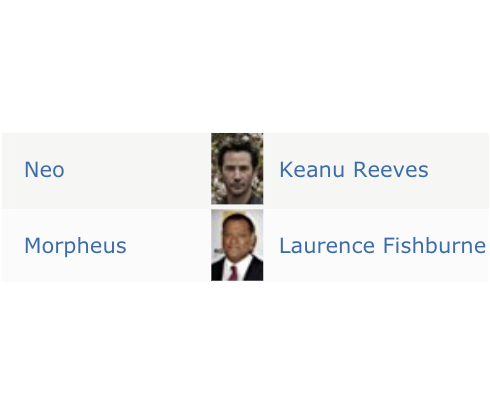}
    \caption{Character list}
  \end{subfigure}
    \begin{subfigure}[c]{0.29\linewidth}
    \includegraphics[height=3cm]{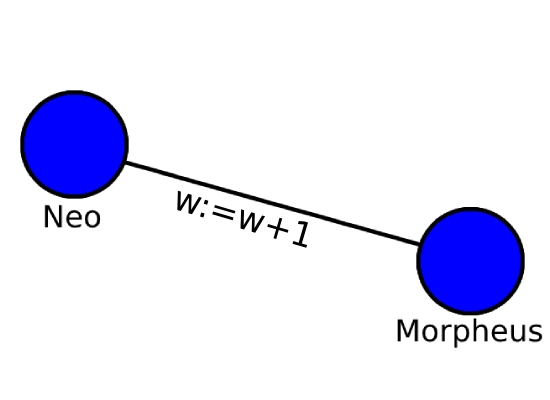}
    \caption{Graph update}
  \end{subfigure}
  \caption{Turning subtitles into a network, step by step:
            (a) perform named entity recognition on the subtitles;
            (b) match the entities to the movie characters; and
            (c) link the characters and increase the edge weight by one.}
  \label{fig:flow}

\end{figure}
ֿ
For a movie with a given subtitle text and a given character list, we constructed the movie's social network using the following steps (see Figure \ref{fig:flow}):

\begin{enumerate}
\item First, we detected when each character appeared in the subtitles. 
To extract the characters from the subtitles we used NER, extracting all the entities which were labeled as a person or an organization.
Additionally, for each entity, we stored the time the entity appeared in the movie.

\item Next, we matched the entities found in the subtitles with the character list.
It worth mentioning that it is not possible to map one-to-one between the characters in the character list and the characters extracted from the subtitle.
% In many cases, character names in the subtitles do not match perfectly to the character list.
For example, in the movie \textit{The Dark Knight}, Bruce Wayne was referred to as ``Bruce Wayne" 3 times, as ``Bruce" 16 times, and as ``Wayne" 20 times.

\item To address the matching problem, we proposed the following mapping heuristic (see Algorithm \ref{alg:match}).
First, we split all the roles into first and last names and linked them to the actor and the character's full name (line 2).
Then, if there was only one character with a certain first or last name (one-to-one match), we linked to the character all its occurrences in the subtitles (lines 3-5).
However, if we had several characters with the same first or last name, we did not always know who was referred to in the text.
For example, in the movie \textit{Back to The Future} there are three characters with the last name McFly; where only "McFly" was mentioned in the text, we could not determine which character was referenced.
Another challenge we encountered was when only part of the character's name was used.
For instance, in the movie \textit{The Godfather}, the main character is Don Vito Corleone, but he was never mentioned once by his full name because he usually was referred to as "Don Corleone."
Moreover, there are other Corleone family members in the movie.
To overcome this challenge, we used \textit{WRatio} to compare strings and match parts of a name to the full name.
Using \textit{WRatio}, we chose the highest matching character that received a score higher than $Threshold$ (line 6).
\SetKw{KwBy}{to}

\begin{algorithm}[H]
 \KwData{PersonName,Roles,Threshold}
 \KwResult{Matched character}
     $Names \gets PersonName.split()$\;
    \ForEach{$N_i \in Names$}{%
        \If{$Roles[N_i].length = 1$}{
           return $Roles[N_i]$\;
           }
        return $MaxWRatio(PersonName,Roles[N_i], Threshold)$
    }
 \caption{Matching entities in the movie subtitles with the characters.}
 \label{alg:match}
\end{algorithm}

\item In fact, we were able to overcome many of these problems by using hearing-impaired subtitles.
In many hearing-impaired subtitles, the name of the speaking character is part of the text.
This property allowed us to avoid most the problems we described earlier and gain additional information.
For instance, the movie \textit{The Matrix} has a scene in which Morpheus calls Neo, and we can  know this only because of the tag [PHONE RINGS].
Afterward, there is an annotation "MORPHEUS:" which tells us that Morpheus is the one calling. 
Without this annotation, we could not know who is on the other end of the line (see Figure~\ref{fig:hear}).

\begin{figure}[ht!]
  \centering
    \includegraphics[width=0.5\linewidth]{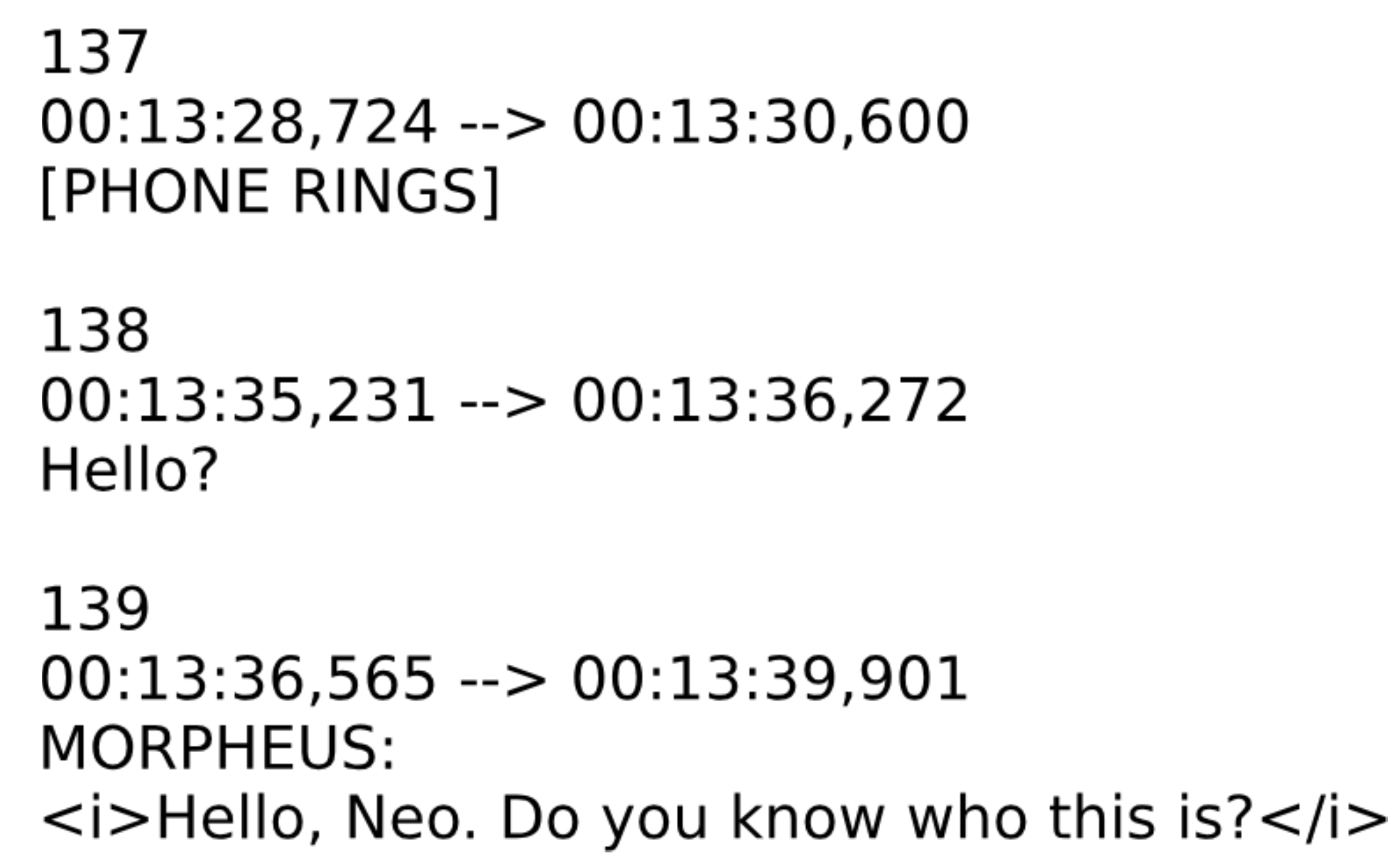}

  \caption{Hearing-impaired subtitles for the movie \textit{The Matrix}.}
  \label{fig:hear}
  \end{figure}
\item Using the matched characters, we created a link between characters $u$ and $v$ if they appeared in the movie in a time interval less than $t$ seconds.
For each such appearance, we increased the weight $w$ between $u$ and $v$ by one.
Since in subtitles we do not have an indication of when each scene begins and ends, we used a heuristic to model the interaction between characters.
We assumed that two characters who appear one after another in a short period of time probably relate.
For example, in Figure~\ref{fig:flow} we have part of the subtitles from the movie \textit{The Matrix}.
Morpheus introduces himself to Neo, and we know that Morpheus and Neo are talking within an interval of 5 seconds.
If $t$ is larger than 5 seconds, we increase the link weight between Morpheus and Neo by one.

\item To reduce the number of false positive edges, we filtered all the edges with weight lower than $w_{min}$.
There were two main reasons for the formation of edges that did not exist in the movie.
The first case was when we matched an entity to the wrong character.
The second case happened when in the interval of \textit{t} seconds there was more than one scene.
These kinds of false positive links add noise to the graph.
Most of these links have a very low weight; hence, filtering edges with weight lower than $w_{min}$ helps remove false positive links.

% We considered a link between character that accruing only few times to be insignificant.
% Sometimes a name only mentioned once in a conversation which just adds noise to the data.
\end{enumerate}

% For film social network we define weighted graph $G$  to be $G:=<V,E,W>$ where set of character in the film represented as a vertex $V$.
% Additionally, $e:=(u,v, w) \in E$ is the set of all edges that represent interaction between two characters $u \in V$ and $v \in V$ and $w$ is the number of co-appearances of $v$ and $u$.
% We define an link between $u$ and $v$ to exist if two character appeared in the film in distance lower than 60 seconds.
% In subtitles we do not have an indication when scene begin and when it ends.
% In order to model the interaction between characters we had to use a heuristic.
% We assumed that two characters who appear one after another in a short amount of time probably relate.
\subsubsection{Evaluations of Constructed Networks}
In addition to constructing movie social networks, we also empirically quantified the quality of these networks.
To evaluate the quality of the constructed networks, we compared them to other publicly available movie network datasets.
Since it is challenging to manually annotate movies, most of the studies only compared their networks to a handful of manually annotated ground truth networks (see Section \ref{sec:rw}).
%As far as we know all the studies that constructed movie social networks compared their networks to handful manually annotated ground truth networks (see Section \ref{sec:rw}).
Comparison to manually annotated networks can only be done at a very modest scale and does not necessarily represent the whole data.

In this study, to the best of our knowledge, we performed the first large-scale, fully-automatic comparison between movie networks.
For the comparison, we used a dataset published in 2018 by Kaminski et al. \cite{kaminski2018moviegalaxies} (denoted as \textit{ScriptNetwork}); this is the only other publicly available movie social network dataset.
The \textit{ScriptNetwork} dataset is based on screenplays and can be considered as much easier content to parse than subtitles.
Screenplays have additional information such as the exact name of the character who speaks in the scene even if this character is unnamed. For example, \textit{freckled kid} is a character in the \textit{X-Men (2000)} screenplay; unnamed characters like \textit{freckled kid} are almost impossible to detect in regular texts like books or subtitles.
Screenplays can be considered very close to the ground truth. However, screenplays sometimes have big differences with the final movie. For instance, in many screenplays, there are missing and even additional characters (see Section \ref{sec:dis}).

To evaluate \textit{Subs2Network}-constructed networks, we performed two types of evaluations:
\begin{itemize}
    \item \textit{Central Character Analysis} - We tested if the most central roles in \textit{Subs2Network} are actually the most central roles in the movie.
As a ground truth, we used the IMDb ranking list similarly to Trans et al. \cite{tran2015cocharnet}.
We tested if the top-5 and top-10 ranked nodes (characters) at \textit{Subs2Network} are the top-5 and top-10 ranked on IMDb.
Additionally, we performed the same test on  networks constructed from screenplays~\cite{kaminski2018moviegalaxies}.
Our motivation behind this experiment was to verify that \textit{Subs2Network's} networks contain the most significant characters in the movie. 

\item \textit{Network Coverage} - We tested if the edges in  \textit{Subs2Network} are the same edges as in other movie networks.
For each movie, we created two sub-graphs containing the characters that exist in both networks.
Then we calculated the edge coverage in the created sub-graphs.
Given two graphs $G$ and $H$, we define the edge coverage as  $Coverage_H(G) = \frac{|E_G \cap E_H| }{|E_H|}$. \break
We calculated $Coverage_{Subs2Network}(\text{ScriptNetwork})$ and   \break $Coverage_{ScriptNetwork}(Subs2Network)$.

\end{itemize}

\subsection{Datasets}
\label{subsec:data}
To evaluate and test our movie social network construction algorithm described above on real-world data, we assembled large-scale datasets of movie subtitles and movie character lists.
In addition, we collected movie character lists from the IMDb (Internet Movie Database) website\footnote{\url{https://www.imdb.com/}} and movie subtitles from 15,540 movies.
Furthermore, we also used data from Bechdel test scores of 4,658 movies.
In the following subsections, we describe in detail the datasets we used.

\subsubsection{IMDb Dataset}
\label{subsec:imdb}
To collect movie and actor data, we used IMDb, which is an online site that contains information related to movies, TV series, video games, etc \cite{PressRoo39:online}.
IMDb data is contributed by users worldwide.
It contains 5,487,394 titles from which 505,380 are full-length movies~\cite{PressRoo63:online}.
In this study, we used the official IMDb dataset.\footnote{\url{https://www.imdb.com/interfaces/}}
From the IMDb dataset, which contains only a subset of the IMDb database, we mainly used movies' titles, crews, and ratings data.

\subsubsection{Subtitle Dataset}

To analyze movies' content, we decided to extract information out of subtitles.
Subtitles are freely and widely available online on numerous sites.
For instance, OpenSubtitles.org\footnote{\url{https://www.opensubtitles.org}} alone hosts more than 500,000 English subtitles \cite{Subtitle14:online} that were manually created by the community.
We collected the subtitles using Subliminal\footnote{\url{https://github.com/Diaoul/subliminal}}, a Python library for searching and downloading subtitles.
Subliminal downloads subtitles from multiple sources, and using an internal scoring method, it decides which subtitles are the best for a specific movie.
Using Subliminal, we downloaded subtitles for 15,540 movies.

\subsubsection{Bechdel Test Dataset}
Bechdel test data is available at Bechdel Test Movie List\footnote{\url{https://bechdeltest.com/}. Note the site uses the Bechdel test variation where women have to have names.},
which is a community-operated website where people can label movies' Bechdel scores.
Using the Bechdel Test Movie List API, we downloaded a dataset that contains 7,871 movies with labeled Bechdel scores, from which only 7,322 are full-length movies.

Even for humans, it is a challenging task to determine if a movie actually passes the Bechdel test; Bechdeltest.com has a comments section where users discuss the scores and their disagreements \cite{agarwal2015key}.
For example, according to Bechdeltest.com, the movie \textit{The Dark Knight Rises} failed the test. However, by taking a closer look at the community comments,\footnote{\url{https://bechdeltest.com/view/3437/the_dark_knight_rises/}} we noticed users arguing regarding the test results, which are hard to determine. 
%if some character was named in the film and if the conversation is about men or not.
% Out of the 7,322 only xxx movies has subtitles available.

% This is also bigger N times than largest analysis of such kind~\cite{agarwal2015key}.

\subsection{Dataset Preprocessing}
The most critical part of building a social network of characters' interaction is mapping correctly between the characters in subtitles and the characters in the character list.
The IMDb character data includes data on even the most minor roles such as a nurse, guard, and thug \#1.
These nameless minor characters are almost impossible to map correctly to their subtitle appearances.
Usually, they just add false positive edges and do not add additional information.

% By examining the IMDb dataset (see Section \ref{subsec:imdb}), we found that the dataset contains over 3.4 million actors while IMDb.com has the data of 38,126,128 actors.
% To address this issue and access the full cast data we used IMDbPY\footnote{\url{https://github.com/alberanid/imdbpy}} a Python package that retrieves data from IMDb website.
% IMDbPY parses IMDb websites to extract the full cast data.

To clean the data from nameless characters, we created a blacklist of minor characters (for a detailed explanation of the blacklist construction process see Section \ref{subsec:blist}).
Additionally, to validate the characters' names we used TMDb (The Movie Database)\footnote{\url{https://www.themoviedb.org}}, another community-built movie database.
For each character, we matched the IMDb and TMDb data by the actor name.
Then, we compared the lengths of the character names and kept the longer one. 

\subsection{Analyzing Movie Social Networks to Identify Gender Bias}

% We wanted to analyze the most popular movies which have a higher impact on the social norms.
% For the dataset, we only collected data of movies which had more than \textit{n} votes on IMDb.
% We collected and fused the data from multiple sources such as IMDb, TMDb, subtitles, etc.
%1,000

\subsubsection{Network Features}
\label{subsec:features}
To study gender bias in movies, we calculated five types of features: vertex features, network features, movie features, gender representation features, and actor features.
Through the study, we analyzed how these features change over time.
Additionally, we used these features to construct machine-learning classifiers.
To create a ground truth for actors' gender, we had to determine whether each actor was male or female.
For most of the characters, we extracted the gender from IMDb similarly to Danescu et al. \cite{danescu2011chameleons}. 
IMDb has an attribute of ``actor'' or ``actress,'' which allowed us to identify gender.
As we mentioned earlier, the IMDb dataset is only partial, so to overcome this issue we used a dataset that maps the first name to the gender.\footnote{\url{http://www.ise.bgu.ac.il/faculty/fire/computationalgenealogy/first_names.html}}
In the rest of this section, we supply the definitions of these features.

\vskip 0.1in

\textbf{Vertex Features:}
For a given $v \in V$, a neighborhood is defined as a set of $v$ friends, $\Gamma(v)$.
Following are the formal definitions of the vertex-based features:
\begin{itemize}
    \item \textit{Total Weight} - the total weight of all the edges, which represents the number of character $v$ appearances in the movie, \linebreak
    $Total_w(v) = \sum_{\{(v, u, w) | ((v, u, w) \in E\}} w$.  
     \item \textit{Closeness Centrality} - the inverse value of the total distance to all the nodes in the graph.
     It is based on the idea that a node closer to other nodes is more central, $ C_c(v) = \frac{1}{\sum_{v \in V}^{} d(v, u)}$ \cite{brandes2005network}, where $d(v, u)$ is the shortest distance between $v$ and $u$.
    \item \textit{Betweenness Centrality} - represents the number of times that a node is a part of the shortest path between two nodes \cite{brandes2005network}.
    A junction (node) that is part of more paths is more central,
    $C_b(v) = \sum_{s,t \in V} \frac{\sigma(s, t|v)}{\sigma(s, t)} \cite{brandes2005network}$,
    where $v \neq s \neq t$, $\sigma(s,t)$ is the number of shortest paths between $s$ and $t$ , and $σ(s,t|v)$ is the number of those paths passing through some node $v$.
    \item \textit{Degree Centrality} - a node that has a higher degree is considered more central, $ C_d(v) = \frac{|\Gamma(v)|}{|V| - 1}$ \cite{brandes2005network}.
    \item \textit{Clustering} - measures link formation between neighboring nodes, $C(v) = \frac{2 T(v)}{|\Gamma(v)|(|\Gamma(v)|-1)}$ \cite{saramaki2007generalizations},
    where $T(v)$ is defined as the number of triangles through vertex $v$ where a triangle is a closed triplet (three vertices that each connect to the other two).
    \item \textit{Pagerank} - a node centrality measure that takes into account the number and the centrality of the nodes pointing to the current node \cite{brandes2005network}.
\end{itemize}

\textbf{Network Features:}
\begin{itemize}
\item \textit{Edge Number} - the number of edges in the network $|E|$.
\item \textit{Vertex Number} - the number of vertices in the network $|V|$.
\item \textit{Number of Cliques} - the number of maximal cliques in the network~\cite{brandes2005network}.
\item \textit{Statistical Network Features} - set of features which are based on the vertex features. 
From these features, we calculate statistical features for the entire network.
We calculate the mean, median, standard deviation, minimum, maximum, first quartile, and third quartile.
\end{itemize}

\textbf{Gender Representation Features:}
\begin{itemize}
\item \textit{Gender Count} - the number of actors of a specific sex in the movie, $G_{c}(G, Sex) = |\{v| v \in V\, gender(v)=Sex\}$.
\item \textit{Triangles with N Women} - the number of triangles that contain N females and 3-N males, where $N \in {1,2,3}$.
\item \textit{Percent of Triangles with N Women} - the percent of triangles that contain N females and 3-N males, where $N \in {1,2,3}$.
\item \textit{Females in Top-10 Roles} - the number of females in top-10 roles ordered by PageRank.
\item \textit{Male Count} - the number of male actors in the movie.
\item \textit{Female Count} - the number of female actors in the movie.
\end{itemize}

\textbf{Movie Features:}
\begin{itemize}
\item \textit{Release Year} - the year when the movie was first aired.
\item \textit{Movie Rating} - the rating the movie has on IMDb.
\item \textit{Runtime} - the movie total runtime in minutes.
\item \textit{Genres} - the movie genre by IMDb.
\item \textit{Number of Votes} - number of votes by which the rating was calculated on IMDb.
\end{itemize}

\textbf{Actor Features:}
\begin{itemize}
\item \textit{Actor Birth Year} - the year the actor was born.
\item \textit{Actor Death Year} - the year the actor died.
\item \textit{Actor Age Filming} - the age of the actor when the movie was released ($Release Year - Actor Birth Year$).
\end{itemize}

\subsubsection{Network Feature Analysis}
% We collected and fused the data from multiple sources such as IMDb, TMDb, subtitles, etc.
To examine the state of the gender gap, in movies generally and by genre in particular, we analyzed only the most popular movies (movies which had more than \textit{n} votes on IMDb).
To answer our first research question -- if there are genres that do not show a gender gap (see Section \ref{sec:int}) -- we calculated vertex and actor features (see Section \ref{subsec:features}) for all the roles.
Next, we split the data by gender and movie genre.
Finally, we utilized a Mann-Whitney U \cite{mann1947test} test on these features to check if there are statistical differences between the male and female roles in different genres.

To study relationships in movies, and to answer our second question regarding what relationships reveal about gender, we calculated all the relationship triangles in the network and grouped them by the number of women in each triangle.
Afterward, we segmented the triangles by genres and how they changed over time.

To investigate the role of centrality by gender, our third research question regardong the centrality of female roles, we calculated PageRank for the nodes in all our movie networks.
We analyzed the number of men and women in the top-10 characters in movies and examined how this number has changed over the years.

\subsubsection{Constructing the Bechdel Test Classifier}
\label{subsec:bechdel}
As we described in Section \ref{sec:rw}, the Bechdel test is used to assess how fairly women are represented in a movie.
The test has three criteria:
\begin{enumerate}
    \item Are there at least two named women in the movie?
    \item Do the women talk to each other?
    \item Do the women talk about something other than men?
\end{enumerate}
These criteria are hierarchical; hence, if a movie passes the last test, it has passed all of the tests.

To train the classifier, we extracted all the network, vertex, and gender representation features (see Section \ref{subsec:features}).
For testing the trained model, we used the \textit{n} newest movies in the Bechdel test dataset. %1000
The rest of the movies were used as the training set.
Additionally, to standardize metrics such as AUC, we evaluated the classifier performance by the \textit{Precision at k} (P@k) metric.
P@k presents how many of the results the classifier is confident it classified correctly.

To answer the fourth research question regarding the fairness of female representation, we analyzed the change in the average probability of a movie passing the Bechdel test over time.
Finally, we analyzed the change over time by genre.

\section{Results}
\label{sec:results}
To analyze the gender gap in the film industry, we analyzed subtitles of movies that had at least 1,000 votes on IMDb.
This resulted in a dataset containing 15,540 movies, which is a dataset 20 times bigger than the largest movie dataset currently available~\cite{kaminski2018moviegalaxies}.

% First, we predicted the actor's gender from movie and vertex features.
% We evaluated our classifier using leave-one-group-out cross-validation and found that we could predict gender by roles significantly better than random.
% Our classifier achieved an AUC of 0.83, and precision and recall of 0.72 and 0.77, respectively.
% We also examined which features were the most important (see Table \ref{tab:imp-gender}); in other words, these features are more biased.

First, we analyzed the gender gap, in general, and by genres, in particular (see Tables \ref{tab:w-test}-\ref{tab:w-test-g4}).
We found that the genre with the largest number of features that are distributed similarly between men and women is Musicals.
In musical movies, 9 out of 10 features distribute similarly; only the clustering coefficient distributes differently between men and women.
In terms of features, \textit{Total Weight} is the feature that distributes  most similarly between the genders, with 9 out of 21 genres distributing the same.
On the other side of the scale, \textit{Age Filming} is the feature that distributes least similarly, with 0 out of 21 genres distributing similarly. 

% \begin{table}[ht!]
% \centering
% \caption{Top-10 most important features in the gender classifier.}
% \label{tab:imp-gender}
% \begin{tabular}{|llll|}
% \hline
% Feature                & Importance & Male Average & Female Average \\   \hline \rowcolor{Gray}
% Actors Age Filming     & 0.20282    & 43           & 35             \\
% Birth Year             & 0.199139   & 1950         & 1960           \\ \rowcolor{Gray}
% Average Movie Rating   & 0.185799   & 6.43         & 6.4            \\ 
% Weighted PageRank      & 0.06171    & 0.15         & 0.13           \\ \rowcolor{Gray}
% Betweenness            & 0.061318   & 0.167        & 0.125          \\
% Action                 & 0.043147   & 0.23         & 0.14           \\ \rowcolor{Gray}
% Release Year           & 0.032128   & 1993         & 1995           \\
% Total Weight           & 0.026612   & 199          & 19             \\ \rowcolor{Gray}
% Degree                 & 0.024525   & 8.34         & 7.62           \\
% Runtime Minutes        & 0.020493   & 109          & 108            \\ \rowcolor{Gray}
% Clustering Coefficient & 0.017633   & 0.565        & 0.6            \\
% PageRank               & 0.01755    & 0.124        & 0.11           \\ \rowcolor{Gray}
% Weighted Betweenness   & 0.016522   & 0.17         & 0.14          \\ \hline
% \end{tabular}
% \end{table}

\begin{figure}[ht!]
  \centering
    \includegraphics[width=1\linewidth]{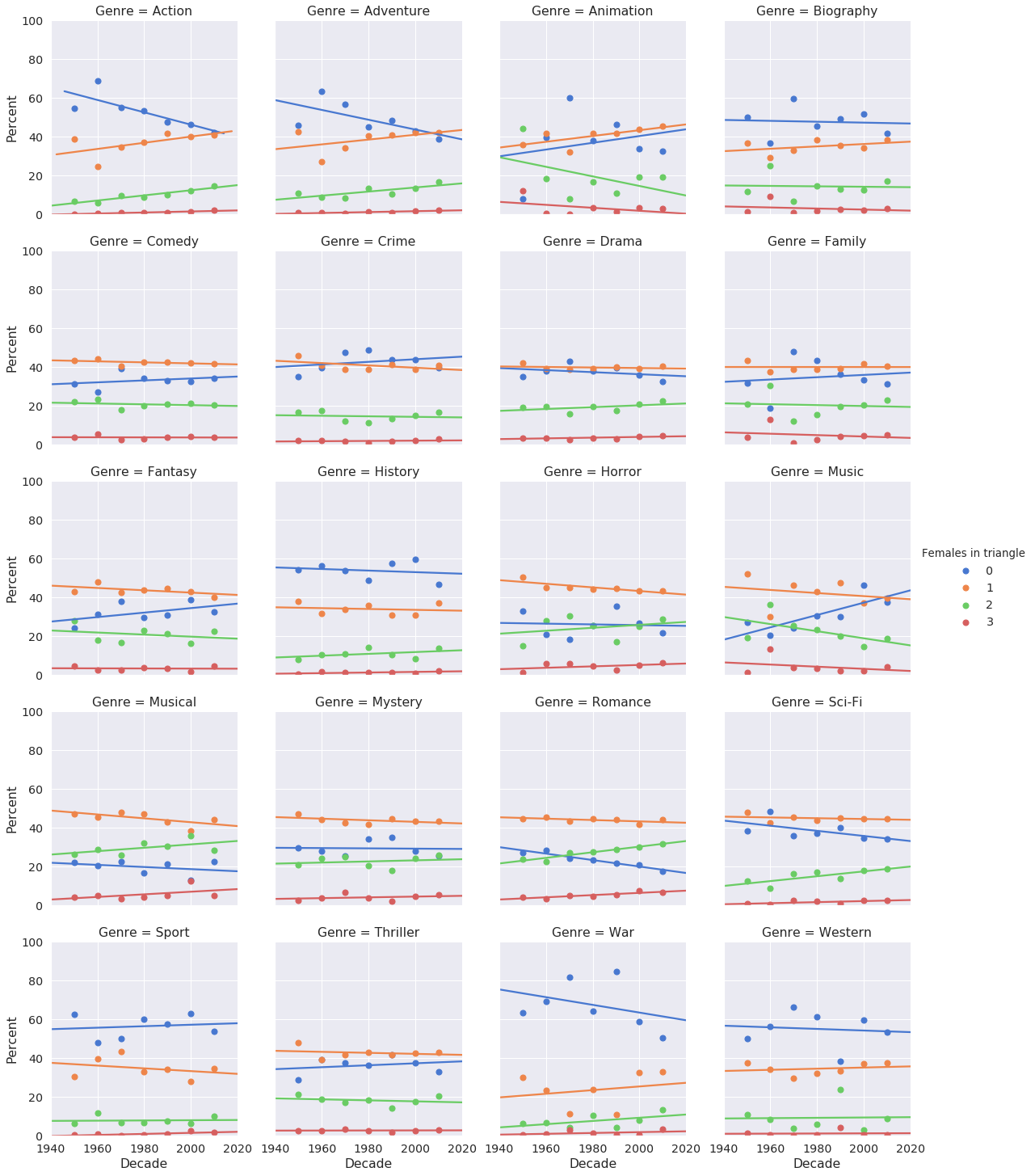}
  \caption{Relationship triangles change over time by different genres.}
  \label{fig:tri-genres}

\end{figure}

Second, to examine relationships among characters, we analyzed relationship triangles in the networks.
We found that most triangles have three men, and triangles with three women are the least common (see Table \ref{tab:tri}).
Out of 21 genres, in 8 genres the most common type of triangle is 3 men (without any women) and in all the others it is 2 men and a woman.
According to the results, Romance is the genre with the most interaction among women and War is the genre where women have the least interaction.
Inspecting the change in the number of triangles over time (see Figure \ref{fig:tri-genres}), we can observe that in many genres there is an equalizing improvement over the years, but there are genres like Sport without a significant change.

\begin{table}[ht!]
\centering
\caption{Relationship triangles in the social network.}
\label{tab:tri}
\begin{tabular}{|l|cccc|}
\hline
Females in triangle & 0       & 1       & 2       & 3      \\ \hline \rowcolor{Gray}
All                 & 40.74\% & 36.56\% & 19.14\% & 3.57\% \\
Action              & 45.85\% & 40.01\% & 12.59\% & 1.55\% \\ \rowcolor{Gray}
Adventure           & 43.36\% & 40.97\% & 13.97\% & 1.70\% \\
Animation           & 34.48\% & 44.36\% & 18.44\% & 2.72\% \\ \rowcolor{Gray}
Biography           & 45.49\% & 36.74\% & 15.09\% & 2.69\% \\
Comedy              & 33.71\% & 41.93\% & 20.53\% & 3.83\% \\ \rowcolor{Gray}
Crime               & 42.54\% & 40.59\% & 14.76\% & 2.10\% \\
Drama               & 35.50\% & 40.01\% & 20.46\% & 4.03\% \\ \rowcolor{Gray}
Family              & 33.04\% & 40.52\% & 21.52\% & 4.93\% \\ 
Fantasy             & 34.24\% & 42.25\% & 20.10\% & 3.41\% \\ \rowcolor{Gray}
Film-Noir           & 35.97\% & 45.59\% & 16.59\% & 1.85\% \\ 
History             & 53.10\% & 34.02\% & 11.30\% & 1.58\% \\ \rowcolor{Gray}
Horror              & 24.71\% & 43.62\% & 26.31\% & 5.36\% \\
Music               & 37.60\% & 40.00\% & 18.78\% & 3.62\% \\ \rowcolor{Gray}
Musical             & 19.59\% & 45.60\% & 29.13\% & 5.68\% \\
Mystery             & 28.78\% & 43.56\% & 23.27\% & 4.39\% \\ \rowcolor{Gray}
Romance             & 21.29\% & 43.61\% & 29.03\% & 6.07\% \\
Sci-Fi              & 35.59\% & 44.71\% & 17.49\% & 2.21\% \\ \rowcolor{Gray}
Sport               & 57.43\% & 32.60\% & 8.27\%  & 1.70\% \\
Thriller            & 36.34\% & 42.65\% & 18.24\% & 2.77\% \\ \rowcolor{Gray}
War                 & 64.24\% & 25.46\% & 8.73\%  & 1.57\% \\ 
Western             & 55.08\% & 35.54\% & 8.50\%  & 0.87\% \\ \hline

\end{tabular}
\end{table}

Third, we analyzed how characters are ranked in terms of centrality (see Tables \ref{tab:deg} and \ref{tab:page}).
We found that among central roles, there are considerably more men than women.
For example, men have about twice the roles that ranked in the top-10 most central roles than women.
In all top-10 most central roles, the female percentage is the same except for the most central role.

% What we did not expected to find is that the representation of women is the same in all top-10 characters in terms of Degree Centrality and PageRank.
% Moreover, we can see that there is not discrimination in the assignment of top roles since in general there are 32.3\% of women roles in films.
% This results indicates the women underrepresented generally and there is no discrimination in terms of centrality there are just less women roles. 

\begin{table}
\caption{The percent of characters by gender, ranked by Degree Centrality in table (a) and PageRank in table (b).}
 \begin{subtable}[b]{0.49\textwidth}
 \centering

\caption{Degree Centrality}
\label{tab:deg}
\begin{tabular}{|lll|}
\hline
Rank & F\%      & M\%      \\ \hline   \rowcolor{Gray}
1    & 28.22\% & 71.78\% \\
2    & 32.19\% & 67.81\% \\ \rowcolor{Gray}
3    & 32.84\% & 67.16\% \\
4    & 32.56\% & 67.44\% \\ \rowcolor{Gray}
5    & 32.54\% & 67.46\% \\
6    & 32.65\% & 67.35\% \\ \rowcolor{Gray}
7    & 32.46\% & 67.54\% \\
8    & 32.16\% & 67.84\% \\ \rowcolor{Gray}
9    & 31.46\% & 68.54\% \\
10   & 32.60\% & 67.40\% \\ \hline
\end{tabular}
\end{subtable}
  \begin{subtable}[b]{0.49\textwidth}
  \centering

\caption{PageRank}
\label{tab:page}
\begin{tabular}{|lll|}
\hline
Rank & F\%      & M\%      \\ \hline  \rowcolor{Gray}
1      & 28.02\% & 71.98\% \\
2      & 32.24\% & 67.76\% \\ \rowcolor{Gray}
3      & 32.84\% & 67.16\% \\
4      & 32.11\% & 67.89\% \\ \rowcolor{Gray}
5      & 32.63\% & 67.37\% \\
6      & 32.81\% & 67.19\% \\ \rowcolor{Gray}
7      & 32.04\% & 67.96\% \\
8      & 32.88\% & 67.12\% \\ \rowcolor{Gray}
9      & 32.14\% & 67.86\% \\
10     & 32.28\% & 67.72\% \\ \hline
\end{tabular}
\end{subtable}
\end{table}

\begin{figure}[ht!]
  \centering
  \begin{subfigure}[t]{0.49\linewidth}
    \includegraphics[width=\linewidth]{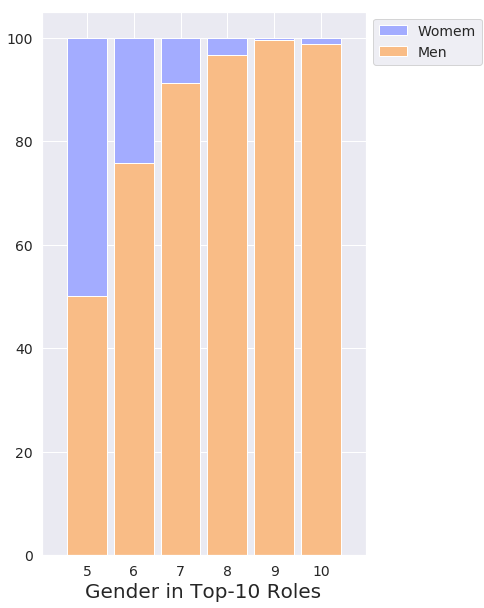}
    \caption{The percentage of movies where out of top-10 role N are of a specific gender.}
  \end{subfigure}
  \begin{subfigure}[t]{0.49\linewidth}
    \includegraphics[width=\linewidth]{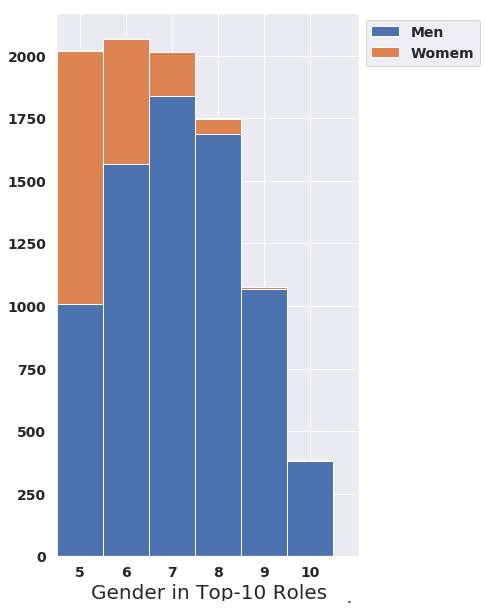}
    \caption{The number of movies where out of top-10 role N are of a specific gender.}
  \end{subfigure}
  \caption{The distribution of movies by  gender of the top-10 most central characters. 
}
\label{fig:dist}

\end{figure}

Fourth, we analyzed the gender composition of the top-10 central roles in movies (see Figure \ref{fig:dist}).
We discovered that most of the movies have more men in central roles than women.
Moreover, from the data, we can observe that there are almost no movies with no men and 10 women in the top-10 roles.
Also, there are a considerable number of movies where the majority of the top-10 most central roles are men. 

\begin{figure}[ht!]
  \centering
    \includegraphics[width=1\linewidth]{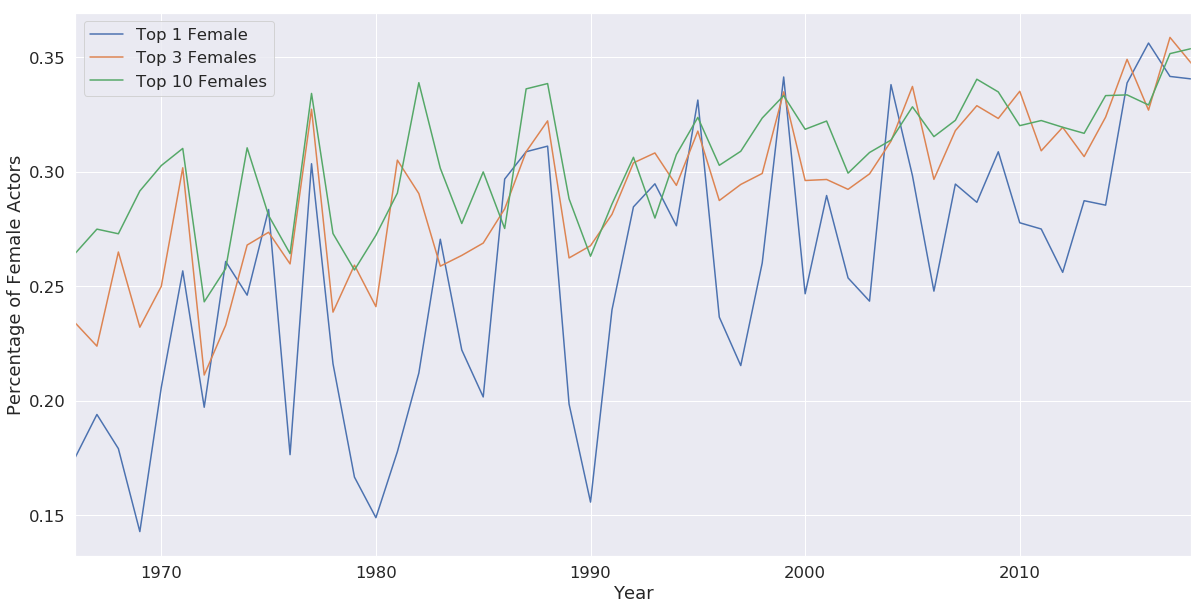}
  \caption{The change in the percentage of women in top 1,3, and 10 most central roles over time.}
  \label{fig:trend}
\end{figure}

Fifth, we wanted to observe how the percentage of women in top 1, 3 and 10 most central roles has evolved over time.
We analyzed the change in this metric over almost a century (see Figure~\ref{fig:trend}).
It can be seen from the network that there is a constant rise in the number of women in top-10 most central roles.

\begin{figure}[ht!]
  \centering
    \includegraphics[width=0.5\linewidth]{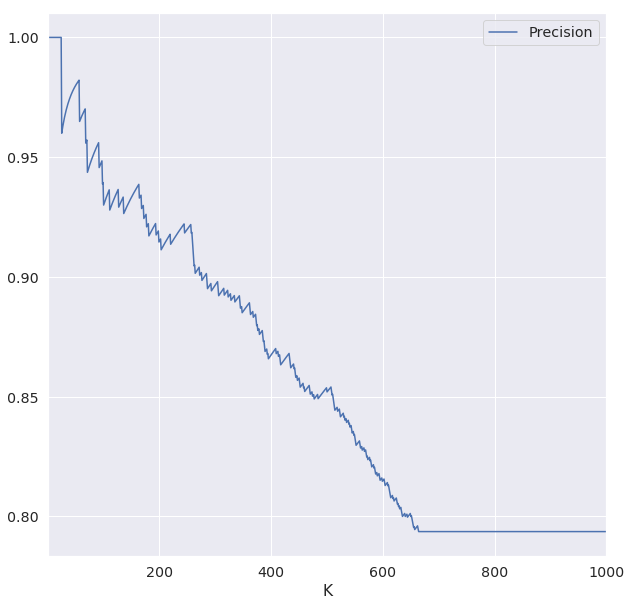}

  \caption{Precision at K of the Bechdel test classifier. }
  \label{fig:precisionk}

\end{figure}

Sixth, to create an automatic classifier that can assess the fairness of female representation in movies, we created the Bechdel test classifier.
Our classifier achieved an AUC of 0.81. 
To check real-world usage of the classifier, we calculated \textit{precision at K} (P@K) (see Figure~\ref{fig:precisionk}).
We can observe that in the first 200 movies in the validation set, we achieved high precision above 0.9.
We also inspected which feature was more important (see Table \ref{tab:imp-bechdel}).
Seven of ten features were triangle-based features.
Moreover, all the features in the table are a subset of the \textit{Gender Representation Features} (see Section~\ref{subsec:features})

\begin{figure}[ht!]
  \centering
  \begin{subfigure}[t]{0.49\linewidth}

    \includegraphics[width=\linewidth]{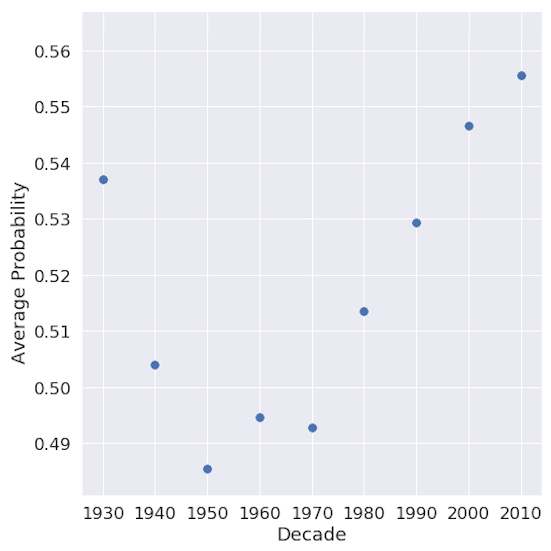}
    \caption{The average probability of passing the Bechdel test by decade.}
  \end{subfigure}
  \begin{subfigure}[t]{0.49\linewidth}
    \includegraphics[width=\linewidth]{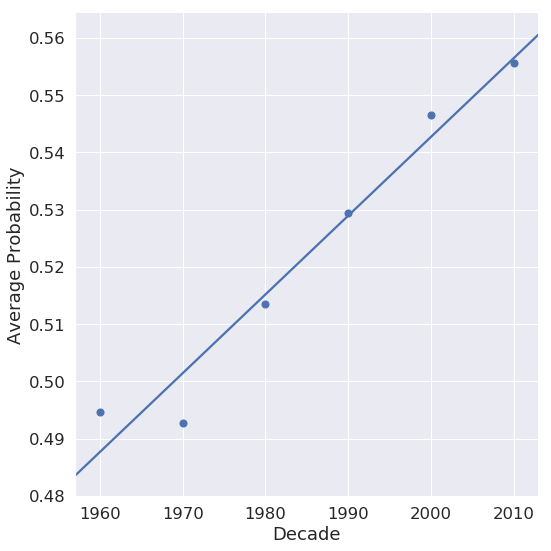}
    \caption{Trend line of probability of passsing the Bechdel test in the past 60 years.}
  \end{subfigure}
  \caption{Probability of a movie passing the Bechdel test by decade. }
  \label{fig:bechdel-decade}
\end{figure}

\begin{figure}[ht!]
  \centering
    \includegraphics[width=1\linewidth]{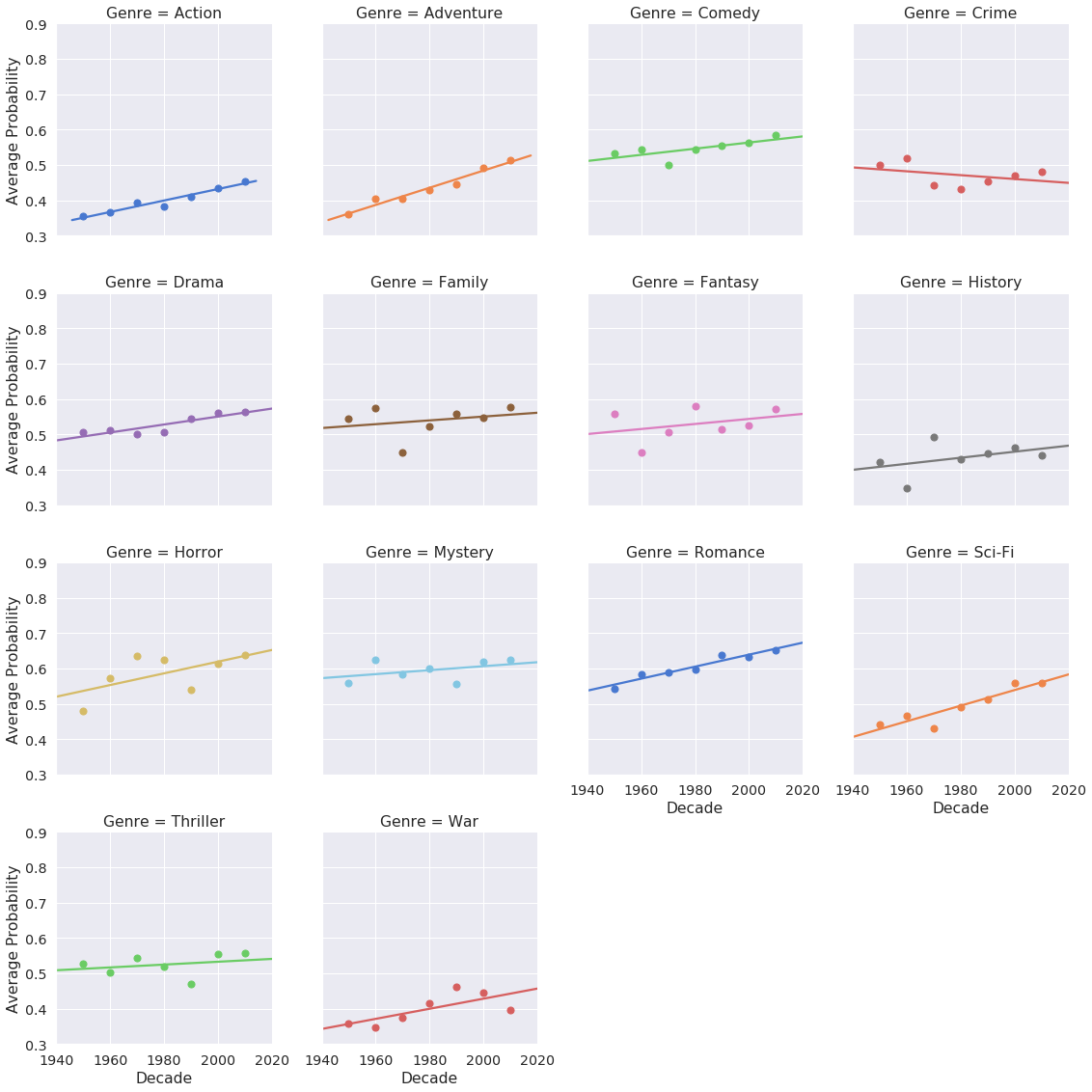}

  \caption{The average probability of a movie passing the Bechdel test by decade and genre. }
  \label{fig:bechdel-decade-genre}

\end{figure}

Next, we trained our automated Bechdel test classifier on all the labeled data and calculated the average probability of the classifier by decade on all the unlabeled data (see Figure \ref{fig:bechdel-decade}).
We can see that there is a trend of growth.
Also, we examined how the probability changed by genres (see Figure \ref{fig:bechdel-decade-genre}).
Comparing our results to Agarwal et al.~\cite{agarwal2015key} (see Table \ref{fig:bechdel-comp}), we found that our classifier performs better than Agarwal's in terms of F1 score.

Finally, we analyzed the quality of the constructed social networks by comparing \textit{Subs2Network} with the \textit{ScriptNetwork}-released networks~\cite{kaminski2018moviegalaxies}.
We observed that the \textit{Subs2Network} dataset contains 628 out of the 773 networks that appear in the \textit{ScriptNetwork} dataset.
In terms of central characters, on average Subs2Network had more central characters than \textit{ScriptNetwork} (see Table \ref{tbl:top-n}); for instance, in the top-10 characters \textit{Subs2Network} matched 6.06 characters while \textit{ScriptNetwork} matched 5.35 characters.
In terms of edge coverage, we found that \textit{Subs2Network} covered 65.4\% of the edges in \textit{ScriptNetwork} networks and \textit{ScriptNetwork} covered 65.1\% of the edges in \textit{Subs2Network} networks.
   
\begin{table}[]
\centering

\begin{tabular}{|l|l|l|}
\hline
                & \textit{ScriptNetwork} & \textit{Subs2Network} \\ \hline \rowcolor{Gray}
 Top-5 & 2.7   &  2.8   \\
Top-10  & 5.35   & 6.06  \\ \hline
\end{tabular}
  \caption{The average number of Top-5,10 most central characters in the movie graph by degree centrality which are also in top-5,10 IMDb most central characters.}
  \label{tbl:top-n}
\end{table}

\begin{table}[]
\centering
\begin{tabular}{|l|lll|lll|}
\hline
               &           & Fail   &       &           & Pass   &      \\ 
               & Precision & Recall & F1    & Precision & Recall & F1   \\ \hline \rowcolor{Gray}
Agrawal et al. & 0.42      & 0.84   & 0.56  & 0.9       & 0.55   & 0.68 \\ 
Current study  & 0.718     & 0.757  & 0.737 & 0.74      & 0.73   & 0.72 \\ \hline
\end{tabular}
  \caption{5-fold cross-validation of the Bechdel test classifier and comparison to the results of Agrawal et al. \cite{agarwal2015key} }
  \label{fig:bechdel-comp}
\end{table}

\begin{table}
\centering
\caption{Top-10 most important features in the gender Bechdel test classifier}
\label{tab:imp-bechdel}
\begin{tabular}{|ll|}
\hline
Feature                         & Importance \\ \hline  \rowcolor{Gray}
Percent of Triangles of 2 Women & 0.157974   \\
Percent of Triangles of 0 Women & 0.14502    \\ \rowcolor{Gray}
Females in Top 10 Roles         & 0.136595   \\
Percent of Triangles of 3 Women & 0.120586   \\ \rowcolor{Gray}
Triangles of 3 Women            & 0.07433    \\
Triangles of 2 Women            & 0.054393   \\ \rowcolor{Gray}
Female Count                    & 0.040251   \\
Triangles of 0 Woman            & 0.030095   \\ \rowcolor{Gray}
Percent of Triangles of 1 Woman & 0.027671   \\
Triangles of 1 Women            & 0.008216   \\
\hline
\end{tabular}
\end{table}

\section{Discussion}
\label{sec:dis}
In this study, we present a method that converts movie subtitles into social networks, and we analyze these networks to study gender disparities in the film industry.
Using this method, we created the largest available corpus of movie character social networks. The method and the corpus are available for use by other researchers to study additional movies and even TV shows, and it has the potential to revolutionize the study of filmed media.

% The ability to classify the gender of characters based on features unrelated to gender demonstrates that there is enough of a difference between male and female roles that they can be distinguished.
% We can see that the most important feature for gender classification is the actor's birth year.
% This result corresponds to previous studies~\cite{glascock2001gender} that have found that female actors, on average, are much younger than male actors.
% Moreover, the Mann-Whitney U test results show that the only feature in films that is consistently different between men and women over all genres is age.

When looking at relationship triangles, we can see that in 77\% of all triangles men are in the majority.
In an equal society, we would expect to find that the number of triangles with three men, with three women, and with two men and two women would be the same.
However, we discovered that, on average, there are 11.4 times more triangles with three men than with three women, and almost twice as many triangles with two men than two women. %1.91
At a deeper level of granularity, we can see a difference in the number of triangles between different movie genres.
The Romance genre has the highest number of triangles that have two and three women.
On the other side of the scale, 90.6\% of triangles in the War genre have a majority of men.
This result makes sense intuitively.
By looking at Figure~\ref{fig:bechdel-dataset}, we can see that genres with a higher percentage of movies that pass the Bechdel test also have a higher percentage of triangles with a majority of women.

In terms of centrality (see Table \ref{tab:deg}), we can see that men have more central roles than women.
We expected to find more females in less central roles, but the percentage of females distributes evenly in the top-10 most central roles.
We believe that these results correspond to the total percentage of women in the dataset, which is 32.3\%.
This number is still lower than the total percentage of female roles in IMDb, which is 37.2\%.

We also analyzed how many roles in a movie's top-10 most central roles are those of women.
Unsurprisingly, there is a dominance of movies with a majority of men.
For instance, all \textit{Lord of the Rings} movies have 10 men in the top-10 roles.
We found only 5 films where all top-10 roles were female, and each of these featured only women  (one of these films is called \textit{The Women}, another movie \textit{Caged} is about a women's prison, and the movie \textit{The Trouble with Angels} is about a girls' school).

We also presented an automated Bechdel test classifier that can help assess the fairness of how women are presented in movies.
% We implemented the model based on the Bechdel test and believe that our model is more accurate than the presented metrics presented in the results section.
We trained our model on data collected from bechdeltest.com, and we have indications that our model is even more accurate than the above presented results.
%We inspected the movies that the classifier was most confident that they pass or fail the test and was wrong.
We found that many movies on bechdeltest.com are misclassified. For example, \textit{The Young Offenders} passes the test on bechdeltest.com (although the site does state this result is `dubious'), but our work classifies it as a fail. The reverse is true for the movie \textit{Never Let Go}.
%For example, the classifier was confident that the movie ``The Young Offenders'' fails the test but it passes, when we inspected user comments we found comments that calims that movie should pass the test.
%On the other side of the scale the classifier confident that the movie ``Never Let Go '' should pass the test but it fails, when we looked on the comments, we saw the following comment: ``This film passes the Bechtel Test 100\%. ...''.
Based on these observations and on the P@K metric, we believe that our classifier can automatically classify movies with high confidence in the classification.  
Moreover, while the Bechdel test is certainly a useful and important test, it fails to account for many parameters such as the centrality of the characters, repression, etc.
Basically, if there is a movie with only two women who appear in one scene and talk about something other than men for 2 seconds, then the movie will pass the traditional Bechdel test. 
However, this is the only test that has data that can be used to train a classifier.
Our classifier partially tackles this problem since it calculates a score of how strongly the movie passes the test. 

To truly solve the issues of the Bechdel test, a better test should be created.
We believe that a good test can be created by comparing the number of interactions according to each gender.
Hence, we propose an interaction test by comparing the total degree of male and female nodes in our movie social networks.
In only 16.7\% of movies do female characters have an equal or higher total degree than male characters.
Moreover, in 55.8\% of movies the total degree of male characters is at least twice as high as female characters.
We think that a good rule of thumb for a movie should be $0.8< \frac{TotalDegree_F}{TotalDegree_M} <1.2$.
Such a test would not be male nor female biased; sadly today only 12\% of all movies pass this test.
In future work, we are planning to perform statistical tests to compare the distributions of the degrees of male and female nodes and present a more accurate test.

We also calculated the average probability of passing the Bechdel test for all the movies in our dataset that do not have a Bechdel test score.
Afterward, we inspected the change in the average probability of movies passing the test over a long period of time and by different genres.
In almost all genres there is a trend of improvement, and there is a correlation between relationship triangles and the Bechdel score. 
%By looking at Table \ref{tab:tri} we can see that in War movies are one of the genres that have the smallest amounts of female triangles.
% We suspect that the high slope is because there a lot what to improve in terms of women representation in War movies.
Looking at Figure \ref{fig:bechdel-decade-genre}, we see that historically war movies have the lowest probability of passing the Bechdel test.

There are many factors that affect our method's accuracy.
The most critical factor is the quality of both the subtitles and the cast information from IMDb.
In movies where the name of the character in the subtitles does not correspond to IMDb data, the actor cannot be linked to a character.
During our study, we stumbled upon subtitles with spelling mistakes and other inconsistencies.
Also, in some movies like superhero movies, we did not know how to link  the different identities of a character with names such as ``Captain America,'' that potentially could be filtered because it looks like a nameless character.
In addition, nameless characters like ``Street Pedestrian'' sometimes eluded our cleaning process.
There is a balance between cleaning the IMDb data too much and not enough.
We observed that more accurate networks were in movies that had hearing-impaired subtitles since they have additional data and are less affected by the NER accuracy.
Some of these limitations will be addressed in future research.
Additionally, there are many different improvements that can done to increase the accuracy of the networks;
for instance, it is possible to use co-reference resolution, train an NER for subtitles, etc.

One of the biggest challenges of this study was to evaluate the quality of the constructed movie networks.
For the evaluation, we compared the networks created by our algorithm with the networks created by screenplay analysis.
%We found on average a 65\% coverage between the edges of the networks.
%Screenplays are easier media to analyze than subtitles, all the character names are explicitly written and there are indications when the scene begins and ends.
Screenplays have easier content to analyze than subtitles, and they contain plenty of structured information, such as character names, scenes, etc.
However, there are also some shortcomings in using screenplays.
First, only a small fraction of movies have screenplays available online.
Currently, the Internet Movie Script Database (IMSDb)\footnote{\url{https://www.imsdb.com/}} has only 1198 scripts, while there are hundreds of thousands of movies' subtitles available online.
Moreover, many publicly available screenplays are drafts and have major differences from the actual movies.
For instance, the \textit{Minority Report} \footnote{\url{https://www.imsdb.com/scripts/Minority-Report.html}} screenplay used by Kaminski et al. is completely different from the movie; almost all the characters' names are different. 
Another example can be found in the \textit{X-Men} (2000) movie where the character \textit{Beast} appears in the screenplay. However, due to over-budget concerns, \textit{Beast} was cut from the movie.
From inspecting screenplays we discovered many additional examples of extra, missing, and renamed characters.
These problems show that comparing subtitles to screenplays is like comparing apples to oranges.  The comparison  indicates that there is a similarity between the networks, but it cannot be used as a precise measure of accuracy.
% screenplays sometimes have big differences with the final movie.

There is no doubt that the presented method is not perfect.
For instance, in the film \textit{Star Wars: Episode VI - Return of the Jedi} (see Figure \ref{fig:infog}), Princess Leia never meets Obi Wan Kenobi. 
Obi Wan Kenobi only talks with Luke about her, which created an edge in the graph.
Nonetheless, from the network evaluation, we learn that the constructed networks represent the movie and have enough correct data to supply insights.
Moreover, many calibrations to the method can be made to improve its accuracy; for instance, we can manually select better subtitles to get more accurate networks. 
Such  calibrations are out of the scope of this study, but in future studies we will explore such options.

Besides utilizing subtitles and screenplays, there are other possible ways to analyze movie content.
The first option is to analyze movie videos as Weng et al. did \cite{weng2009rolenet}.
The problem with video analysis is that it is an expensive process which requires high computational power, especially when the plan is to analyze thousands of full-length movies.
Moreover, most movies are copyrighted and not freely available online.
The second option is to use speech recognition to extract information, which is what Park et al. did.~\cite{park2012social}.
However, this option has similar drawbacks.
%We decided to use movie subtitles for this study since subtitles are free and widely available.

\section{Conclusions}
\label{sec:con}
Data science can provide great insights into many problems, including the gender gap in movies. In this work, we created a massive dataset of movie character interactions to present the largest-to-date social network analysis of gender disparities in the film industry.
We constructed this dataset by fusing data from multiple sources, and then we  analyzed the movie gender gap by examining multiple parameters over the past century.

Our results demonstrate that a gender gap remains in nearly all genres of the film industry.
For instance, 3.5 times more relationship triangles in movies have a majority of men.
In terms of top-10 most central movie roles, again there is a majority of men.
However, we also saw an improvement in equality over the years.
Today, women have more important movie roles than in the past, and our Bechdel test classifier quantifies this improvement over time by calculating a movie's overall score.
In a future study, we plan to analyze TV series, actors' careers, and directors' careers in a similar in-depth manner.
We also plan to implement the tests that were proposed in~\cite{Creating26:online} as well as develop new tests to gain further insight into how genders are represented in the film industry.

ֿ\section{Data and Code Availability}
This study is reproducible research. Therefore, the anonymous versions of the social network datasets and the study's code, including implementation, are available on the project's website\footnote{\url{http://data4good.io/dataset.html\#Movie-Dynamics}} and repository\footnote{\url{https://github.com/data4goodlab/subs2network}}.

\section{Acknowledgements}
% We would like to thank Carol Teegarden for editing and proofreading this article to completion. Also, we thank Sean McNaughton, Mandy Henner, Sergey Korotchenko, and Ariel Plotkin for their help.

 We would like to thank Carol Teegarden for editing and proofreading this article to completion. Also, we thank Sean McNaughton Mandy Henner, Sergey Korotchenko, and Ariel Plotkin for their help.

% \section{Highlights}
% \begin{itemize}
%     \item Largest available dataset of movie social networks (15,540 networks). 
%     \item In-depth gender gap analysis over the past century.
%     \item Detection of improvement trend in female representation.
%     \item Development of automated Bechdel test classifier.
%     \item Recommendation of new interaction test as a better alternative to the Bechdel test.

% \end{itemize}
%\end{document}  % This is where a 'short' article might terminate

\bibliographystyle{abbrv}
\bibliography{sample-bibliography}
\appendix
\section{Appendix}
\subsection{Character Blacklist Construction}
\label{subsec:blist}
\begin{enumerate}
    \item First, we initialized a list of all characters from the IMDb dataset (see Section \ref{subsec:data}.
    \item To remove all named characters, we downloaded the U.S Social Security
baby name dataset\footnote{\url{https://www.ssa.gov/oact/babynames/names.zip}} and the U.S Census surname dataset.\footnote{\url{https://www2.census.gov/topics/genealogy/2010surnames/names.zip}}
    We removed all the characters whose names matched the names in these datasets.
    \item Next, we grouped all the characters by name and actor, and we filtered all
characters portrayed by the same actor in more than one film.
    \item Afterward, we aggregated the remaining characters by their names and
counted the number of appearances and the average order of appearance.
We removed all the character names that on average placed in the first
three positions in the cast list. The IMDb cast order mostly represents the importance of the characters, but we noticed some anomalies in the ordering.
In this study, we assumed that the first three roles on IMDb are the main
characters.
    \item Finally, we removed all character names that appeared only once. Generic
character names like Mom, Dad, Policeman, etc., appear in multiple unrelated movies. A character name that appears only in one film has a higher
probability of being an important character.

    \item All the remaining character names were blacklisted.

\end{enumerate}

\subsection{Figures and Tables}

\begin{figure}[ht!]
  \centering
    \includegraphics[height=1\textheight]{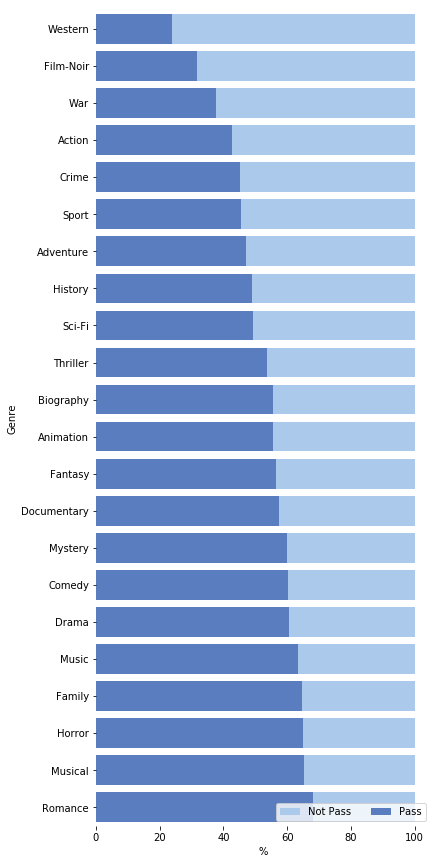}

  \caption{The distribution of the label in Bechdel dataset .}
  \label{fig:bechdel-dataset}

\end{figure}

\begin{table}[h!]
\begin{tabular}{|lllll|}
\hline
Feature            & U          & Median(M) & Median(F) & p-value     \\ \hline  \rowcolor{Gray}
Age Filming        & 44918391.50 & 42.00     & 33.00     & 0   \\ 
Betweenness        & 62658633.50 & 0.08      & 0.05      & 2.96E-54 \\ \rowcolor{Gray}
Closeness          & 64487986.00 & 0.75      & 0.72      & 6.31E-34 \\
Clustering         & 64536999.50 & 0.56      & 0.62      & 1.62E-33 \\ \rowcolor{Gray}
Degree             & 66190111.50 & 7.00      & 7.00      & 1.17E-19 \\ 
Degree Centrality  & 64208427.50 & 0.69      & 0.63      & 1.12E-36 \\ \rowcolor{Gray}
Pagerank           & 64940724.50 & 0.11      & 0.11      & 1.18E-29 \\ 
Weighted Betweenness & 66888718.00 & 0.11      & 0.09      & 4.06E-15 \\   \rowcolor{Gray}
Weighted Pagerank  & 63235436.00 & 0.14      & 0.12      & 4.99E-47 \\
Total Weight       & 67797466.00 & 104.00    & 95.00     & 7.47E-10 \\
\hline
\end{tabular}
\caption{Mann–Whitney U test between men and women.}
\label{tab:w-test}
\end{table}

\begin{table}[]
\scriptsize

\begin{tabular}{|llllll|}
\hline
Genre     & Feature            & U         & Median(M) & Median(F) & p-value     \\ \hline  \rowcolor{Gray}
Action    & Age Filming        & 1141522.5  & 43          & 32          & 1.06E-155   \\
Action    & Betweenness        & 1857629    & 0.090891053 & 0.036363636 & 1.41E-26    \\ \rowcolor{Gray}
Action    & Betweenness Weight & 2043730    & 0.111111111 & 0.072727273 & 5.24E-11    \\
Action    & Closeness          & 1941924.5  & 0.75        & 0.692307692 & 1.78E-18    \\ \rowcolor{Gray}
Action    & Clustering         & 2037316.5  & 0.533333333 & 0.636363636 & 2.53E-11    \\
Action    & Degree             & 2077375    & 7           & 6           & 6.82E-09    \\ \rowcolor{Gray}
Action    & Degree Centrality  & 1925104    & 0.666666667 & 0.571428571 & 6.08E-20    \\
Action    & Pagerank           & 1931693    & 0.113907479 & 0.096573682 & 2.44E-19    \\ \rowcolor{Gray}
Action    & Pagerank Weight    & 1888486.5  & 0.140696271 & 0.104831315 & 2.57E-23    \\
Action    & Total Weight       & 2138621    & 91          & 76          & 8.66E-06    \\ \rowcolor{Gray}
Adventure & Age Filming        & 668865.5   & 43          & 33          & 4.97E-93    \\
Adventure & Betweenness        & 1018363    & 0.072065437 & 0.042779044 & 2.61E-14    \\ \rowcolor{Gray}
Adventure & Betweenness Weight & 1146136.5  & 0.1         & 0.078472222 & 0.002577084 \\
Adventure & Closeness          & 1056896.5  & 0.75        & 0.714285714 & 6.02E-10    \\ \rowcolor{Gray}
Adventure & Clustering         & 1082650    & 0.580882353 & 0.65447861  & 1.43E-07    \\
Adventure & Degree             & 1114523    & 7           & 7           & 3.85E-05    \\ \rowcolor{Gray}
Adventure & Degree Centrality  & 1051316    & 0.666666667 & 0.6         & 1.63E-10    \\
Adventure & Pagerank           & 1073820    & 0.109176206 & 0.098911737 & 2.53E-08    \\ \rowcolor{Gray}
Adventure & Pagerank Weight    & 1015396.5  & 0.136392253 & 0.107955414 & 1.38E-14    \\
Adventure & Total Weight       & 1132409.5  & 105         & 90          & 0.000513202 \\ \rowcolor{Gray}
Animation & Age Filming        & 26773      & 46          & 38          & 1.35E-15    \\
Animation & Betweenness        & 39955      & 0.055810878 & 0.033617725 & 0.029420016 \\ \rowcolor{Gray}
Animation & Betweenness Weight & 40912.5    & 0.071428571 & 0.095238095 & 0.072545277 \\
Animation & Closeness          & 38732.5    & 0.75        & 0.714285714 & 0.007374173 \\ \rowcolor{Gray}
Animation & Clustering         & 42076.5    & 0.636363636 & 0.666666667 & 0.181192059 \\
Animation & Degree             & 39161      & 7           & 6           & 0.012322396 \\ \rowcolor{Gray}
Animation & Degree Centrality  & 38499      & 0.666666667 & 0.6         & 0.005459331 \\
Animation & Pagerank           & 41147.5    & 0.106196092 & 0.100198011 & 0.091332701 \\ \rowcolor{Gray}
Animation & Pagerank Weight    & 37378      & 0.129233386 & 0.101397903 & 0.001124444 \\
Animation & Total Weight       & 35999      & 112         & 69.5        & 0.000114125 \\ \rowcolor{Gray}
Biography & Age Filming        & 225818.5   & 41          & 35          & 1.21E-20    \\
Biography & Betweenness        & 278486     & 0.074104894 & 0.049692308 & 0.000146971 \\ \rowcolor{Gray}
Biography & Betweenness Weight & 287077     & 0.111111111 & 0.089404919 & 0.003482307 \\
Biography & Closeness          & 286549     & 0.722222222 & 0.7         & 0.003033074 \\ \rowcolor{Gray}
Biography & Clustering         & 289311.5   & 0.5         & 0.555555556 & 0.007117802 \\
Biography & Degree             & 274713     & 9           & 7           & 2.98E-05    \\ \rowcolor{Gray}
Biography & Degree Centrality  & 287900     & 0.625       & 0.578947368 & 0.004666262 \\
Biography & Pagerank           & 302390     & 0.09683747  & 0.095406148 & 0.147192227 \\ \rowcolor{Gray}
Biography & Pagerank Weight    & 290355     & 0.115042243 & 0.10650003  & 0.009751414 \\
Biography & Total Weight       & 285146.5   & 124         & 102.5       & 0.001908454 \\ \rowcolor{Gray}
Comedy    & Age Filming        & 7323590.5  & 41          & 34          & 8.79E-160   \\
Comedy    & Betweenness        & 9596319    & 0.088888889 & 0.051340073 & 1.99E-23    \\ \rowcolor{Gray}
Comedy    & Betweenness Weight & 10212374.5 & 0.115384615 & 0.091666667 & 5.99E-08    \\
Comedy    & Closeness          & 9649651.5  & 0.765686275 & 0.722222222 & 1.23E-21    \\ \rowcolor{Gray}
Comedy    & Clustering         & 9913519.5  & 0.555555556 & 0.611111111 & 2.94E-14    \\
Comedy    & Degree             & 10274875   & 8           & 7           & 7.87E-07    \\ \rowcolor{Gray}
Comedy    & Degree Centrality  & 9604776    & 0.7         & 0.625       & 4.62E-23    \\
Comedy    & Pagerank           & 9586716    & 0.115125241 & 0.103417249 & 1.32E-23    \\ \rowcolor{Gray}
Comedy    & Pagerank Weight    & 9358473.5  & 0.147601675 & 0.120580764 & 1.08E-31    \\
Comedy    & Total Weight       & 10277578.5 & 121         & 109.5       & 9.22E-07    \\ \rowcolor{Gray}
Crime     & Age Filming        & 1608067.5  & 42          & 33          & 3.36E-121   \\
Crime     & Betweenness        & 2317850    & 0.084900202 & 0.041666667 & 1.57E-20    \\ \rowcolor{Gray}
Crime     & Betweenness Weight & 2534976    & 0.107142857 & 0.080645855 & 5.32E-07    \\
Crime     & Closeness          & 2419778.5  & 0.75        & 0.714285714 & 4.29E-13    \\ \rowcolor{Gray}
Crime     & Clustering         & 2381780.5  & 0.545454545 & 0.658241758 & 1.19E-15    \\
Crime     & Degree             & 2500022.5  & 7           & 7           & 1.38E-08    \\ \rowcolor{Gray}
Crime     & Degree Centrality  & 2403978.5  & 0.666666667 & 0.6         & 4.08E-14    \\
Crime     & Pagerank           & 2413828    & 0.112332161 & 0.099630832 & 1.86E-13    \\ \rowcolor{Gray}
Crime     & Pagerank Weight    & 2309188    & 0.140377213 & 0.107617666 & 4.10E-21    \\
Crime     & Total Weight       & 2574489.5  & 108         & 95          & 2.52E-05 \\ \hline 
\end{tabular}
\caption{Mann–Whitney U test between men and women and by movie genre.}
\label{tab:w-test-g1}
\end{table}
\begin{table}[]
\scriptsize
\begin{tabular}{|llllll|}
\hline
Genre     & Feature            & U         & Median(M) & Median(F) & p-value     \\ \hline  \rowcolor{Gray}

Family    & Age Filming        & 109501   & 43          & 35          & 3.19E-17    \\
Family    & Betweenness        & 135028.5 & 0.069483182 & 0.038628118 & 0.00010327  \\ \rowcolor{Gray}
Family    & Betweenness Weight & 152510.5 & 0.083333333 & 0.091666667 & 0.307231891 \\
Family    & Closeness          & 142036   & 0.769230769 & 0.75        & 0.007884111 \\ \rowcolor{Gray}
Family    & Clustering         & 137921.5 & 0.6         & 0.666666667 & 0.000755546 \\
Family    & Degree             & 150287   & 8           & 7           & 0.182202283 \\ \rowcolor{Gray}
Family    & Degree Centrality  & 141678   & 0.7         & 0.666666667 & 0.006574201 \\
Family    & Pagerank           & 132781   & 0.107232427 & 0.092132022 & 2.04E-05    \\ \rowcolor{Gray}
Family    & Pagerank Weight    & 127825   & 0.129170457 & 0.10137092  & 2.74E-07    \\
Family    & Total Weight       & 147673   & 126         & 115         & 0.083404211 \\ \rowcolor{Gray}
Fantasy   & Age Filming        & 145618   & 42          & 32          & 4.16E-37    \\
Fantasy   & Betweenness        & 210166.5 & 0.088690476 & 0.044191919 & 6.54E-06    \\ \rowcolor{Gray}
Fantasy   & Betweenness Weight & 216448.5 & 0.116666667 & 0.075757576 & 0.000188026 \\
Fantasy   & Closeness          & 215782.5 & 0.75        & 0.714285714 & 0.000150319 \\ \rowcolor{Gray}
Fantasy   & Clustering         & 224803   & 0.533333333 & 0.636363636 & 0.007070388 \\
Fantasy   & Degree             & 219762.5 & 7           & 7           & 0.000947644 \\ \rowcolor{Gray}
Fantasy   & Degree Centrality  & 212925.5 & 0.6875      & 0.6         & 3.39E-05    \\
Fantasy   & Pagerank           & 218093.5 & 0.114703841 & 0.103376407 & 0.000460687 \\ \rowcolor{Gray}
Fantasy   & Pagerank Weight    & 213766   & 0.140384021 & 0.113793873 & 5.39E-05    \\
Fantasy   & Total Weight       & 225173   & 104         & 93          & 0.008206218 \\ \rowcolor{Gray}
Film-Noir & Age Filming        & 7954     & 39          & 30          & 1.95E-26    \\
Film-Noir & Betweenness        & 18890.5  & 0.050614478 & 0.034864872 & 0.029698627 \\ \rowcolor{Gray}
Film-Noir & Betweenness Weight & 19222.5  & 0.044444444 & 0.090013228 & 0.047625203 \\
Film-Noir & Closeness          & 19247    & 0.875       & 0.851648352 & 0.053007773 \\ \rowcolor{Gray}
Film-Noir & Clustering         & 18698    & 0.722222222 & 0.780952381 & 0.020820651 \\
Film-Noir & Degree             & 17960.5  & 7           & 6           & 0.004192441 \\ \rowcolor{Gray}
Film-Noir & Degree Centrality  & 19209    & 0.857142857 & 0.825757576 & 0.049777976 \\
Film-Noir & Pagerank           & 21070.5  & 0.13009992  & 0.130010016 & 0.442810363 \\ \rowcolor{Gray}
Film-Noir & Pagerank Weight    & 18199    & 0.174634239 & 0.147592919 & 0.007634682 \\
Film-Noir & Total Weight       & 18142.5  & 184         & 137         & 0.00673959  \\ \rowcolor{Gray}
History   & Age Filming        & 55474    & 41          & 34          & 1.05E-20    \\
History   & Betweenness        & 79958.5  & 0.078484848 & 0.043158605 & 0.001836765 \\ \rowcolor{Gray}
History   & Betweenness Weight & 79259.5  & 0.102941176 & 0.053571429 & 0.001008279 \\
History   & Closeness          & 84682.5  & 0.708333333 & 0.689903846 & 0.047765001 \\ \rowcolor{Gray}
History   & Clustering         & 83098.5  & 0.522875817 & 0.6         & 0.018700742 \\
History   & Degree             & 84944    & 8           & 7           & 0.054633765 \\ \rowcolor{Gray}
History   & Degree Centrality  & 84356.5  & 0.6         & 0.571428571 & 0.039914883 \\
History   & Pagerank           & 84616.5  & 0.09843372  & 0.08690905  & 0.046129716 \\ \rowcolor{Gray}
History   & Pagerank Weight    & 82926    & 0.115658997 & 0.097958953 & 0.01688756  \\
History   & Total Weight       & 86931    & 82          & 83          & 0.139541929 \\ \rowcolor{Gray}
Horror    & Age Filming        & 232696.5 & 43          & 33          & 5.66E-51    \\
Horror    & Betweenness        & 391571.5 & 0.063369963 & 0.048636364 & 0.202004507 \\ \rowcolor{Gray}
Horror    & Betweenness Weight & 376206.5 & 0.091911765 & 0.066666667 & 0.012800634 \\
Horror    & Closeness          & 400103   & 0.75        & 0.764705882 & 0.472054474 \\ \rowcolor{Gray}
Horror    & Clustering         & 388844.5 & 0.642857143 & 0.666666667 & 0.141097096 \\
Horror    & Degree             & 392754   & 6           & 6           & 0.233654523 \\ \rowcolor{Gray}
Horror    & Degree Centrality  & 400763.5 & 0.684210526 & 0.7         & 0.495519182 \\
Horror    & Pagerank           & 393557   & 0.120876857 & 0.119215309 & 0.257037519 \\ \rowcolor{Gray}
Horror    & Pagerank Weight    & 392331.5 & 0.142857143 & 0.141634331 & 0.223161186 \\
Horror    & Total Weight       & 376062   & 83          & 95          & 0.013573873 \\ \rowcolor{Gray}
Music     & Age Filming        & 36852    & 38          & 31          & 6.12E-09    \\
Music     & Betweenness        & 43732.5  & 0.105603656 & 0.067424242 & 0.00281239  \\ \rowcolor{Gray}
Music     & Betweenness Weight & 46984.5  & 0.137362637 & 0.104166667 & 0.083124805 \\
Music     & Closeness          & 42434.5  & 0.761904762 & 0.708333333 & 0.00045222  \\ \rowcolor{Gray}
Music     & Clustering         & 48215    & 0.523809524 & 0.558080808 & 0.196817044 \\
Music     & Degree             & 47201    & 7           & 7           & 0.098921715 \\ \rowcolor{Gray}
Music     & Degree Centrality  & 42609    & 0.7         & 0.6         & 0.000588954 \\
Music     & Pagerank           & 43775    & 0.120154842 & 0.10723231  & 0.003033951 \\ \rowcolor{Gray}
Music     & Pagerank Weight    & 43148    & 0.151410784 & 0.119035137 & 0.001300769 \\
Music     & Total Weight       & 46491    & 107         & 94          & 0.056285977 \\ \hline
\end{tabular}
\caption{Mann–Whitney U test between men and women and by movie genre.}
\label{tab:w-test-g2}
\end{table}

\begin{table}[]
\scriptsize
\begin{tabular}{|llllll|}
\hline
Genre     & Feature            & U         & Median(M) & Median(F) & p-value     \\ \hline \rowcolor{Gray}
Musical  & Age Filming        & 18923.5   & 39          & 30          & 6.24E-17    \\
Musical  & Betweenness        & 31309     & 0.044642857 & 0.054444444 & 0.147010043 \\ \rowcolor{Gray}
Musical  & Betweenness Weight & 31851.5   & 0.066666667 & 0.079861111 & 0.22699751  \\
Musical  & Closeness          & 31085     & 0.8         & 0.833333333 & 0.117765111 \\ \rowcolor{Gray}
Musical  & Clustering         & 31953.5   & 0.690909091 & 0.691666667 & 0.251068339 \\
Musical  & Degree             & 30671.5   & 6           & 7           & 0.076823319 \\ \rowcolor{Gray}
Musical  & Degree Centrality  & 31199     & 0.75        & 0.80625     & 0.131591811 \\
Musical  & Pagerank           & 31495     & 0.123660566 & 0.123992966 & 0.174486394 \\ \rowcolor{Gray}
Musical  & Pagerank Weight    & 32155     & 0.155498559 & 0.161792306 & 0.290705413 \\
Musical  & Total Weight       & 31556     & 139         & 152.5       & 0.183798689 \\ \rowcolor{Gray}
Mystery  & Age Filming        & 299382    & 44          & 33          & 4.66E-66    \\
Mystery  & Betweenness        & 511460.5  & 0.064229055 & 0.049481074 & 0.025836206 \\ \rowcolor{Gray}
Mystery  & Betweenness Weight & 534156    & 0.081818182 & 0.089285714 & 0.376526415 \\
Mystery  & Closeness          & 512182.5  & 0.760952381 & 0.75        & 0.029565805 \\ \rowcolor{Gray}
Mystery  & Clustering         & 502014.5  & 0.6         & 0.666666667 & 0.004400255 \\
Mystery  & Degree             & 503329.5  & 7           & 7           & 0.005779962 \\ \rowcolor{Gray}
Mystery  & Degree Centrality  & 510327.5  & 0.692307692 & 0.666666667 & 0.021679249 \\
Mystery  & Pagerank           & 535135.5  & 0.111685765 & 0.110774791 & 0.404877623 \\ \rowcolor{Gray}
Mystery  & Pagerank Weight    & 526377.5  & 0.135672521 & 0.127442474 & 0.192731844 \\
Mystery  & Total Weight       & 519828    & 105.5       & 100.5       & 0.090631078 \\ \rowcolor{Gray}
Romance  & Age Filming        & 2419166   & 39          & 32          & 8.11E-94    \\
Romance  & Betweenness        & 3387209.5 & 0.083333333 & 0.070011338 & 0.000238122 \\ \rowcolor{Gray}
Romance  & Betweenness Weight & 3509329.5 & 0.109090909 & 0.100608466 & 0.089109724 \\
Romance  & Closeness          & 3321353   & 0.777777778 & 0.75        & 1.68E-06    \\ \rowcolor{Gray}
Romance  & Clustering         & 3481520   & 0.582117882 & 0.6         & 0.033891956 \\
Romance  & Degree             & 3575369   & 7           & 7           & 0.431392273 \\ \rowcolor{Gray}
Romance  & Degree Centrality  & 3322731   & 0.722222222 & 0.666666667 & 1.89E-06    \\
Romance  & Pagerank           & 3346711.5 & 0.121533126 & 0.114621794 & 1.37E-05    \\ \rowcolor{Gray}
Romance  & Pagerank Weight    & 3310323.5 & 0.15475242  & 0.140722332 & 6.71E-07    \\
Romance  & Total Weight       & 3539603.5 & 112         & 113         & 0.211472324 \\ \rowcolor{Gray}
Sci-Fi   & Age Filming        & 87562     & 42          & 32          & 4.15E-51    \\
Sci-Fi   & Betweenness        & 159099.5  & 0.078979592 & 0.045436508 & 3.96E-05    \\ \rowcolor{Gray}
Sci-Fi   & Betweenness Weight & 168266.5  & 0.100088183 & 0.071428571 & 0.00548229  \\
Sci-Fi   & Closeness          & 166190.5  & 0.769230769 & 0.729020979 & 0.002240595 \\ \rowcolor{Gray}
Sci-Fi   & Clustering         & 163459    & 0.583333333 & 0.666666667 & 0.000543829 \\
Sci-Fi   & Degree             & 168407.5  & 7           & 7           & 0.006140502 \\ \rowcolor{Gray}
Sci-Fi   & Degree Centrality  & 165481.5  & 0.705882353 & 0.642857143 & 0.001579692 \\
Sci-Fi   & Pagerank           & 170767    & 0.115872323 & 0.106716646 & 0.016493957 \\ \rowcolor{Gray}
Sci-Fi   & Pagerank Weight    & 166915    & 0.144101358 & 0.122172161 & 0.003192407 \\
Sci-Fi   & Total Weight       & 177776    & 103         & 97.5        & 0.146996378 \\ \rowcolor{Gray}
Sport    & Age Filming        & 18836.5   & 41          & 34          & 6.96E-11    \\
Sport    & Betweenness        & 19327.5   & 0.088212251 & 0.016666667 & 4.07E-10    \\ \rowcolor{Gray}
Sport    & Betweenness Weight & 23440     & 0.116506785 & 0.049206349 & 0.000113214 \\
Sport    & Closeness          & 20763     & 0.75        & 0.647058824 & 6.87E-08    \\ \rowcolor{Gray}
Sport    & Clustering         & 20805     & 0.514928699 & 0.666666667 & 7.52E-08    \\
Sport    & Degree             & 22075.5   & 9           & 7           & 3.57E-06    \\ \rowcolor{Gray}
Sport    & Degree Centrality  & 20514     & 0.666666667 & 0.5         & 3.02E-08    \\
Sport    & Pagerank           & 21340.5   & 0.096916494 & 0.071477478 & 4.30E-07    \\ \rowcolor{Gray}
Sport    & Pagerank Weight    & 19037     & 0.124949309 & 0.065430363 & 1.54E-10    \\
Sport    & Total Weight       & 21747     & 135         & 83          & 1.44E-06    \\ \rowcolor{Gray}
Thriller & Age Filming        & 821514    & 43          & 33          & 4.09E-113   \\
Thriller & Betweenness        & 1363220   & 0.082063492 & 0.05042735  & 9.61E-08    \\ \rowcolor{Gray}
Thriller & Betweenness Weight & 1435399   & 0.10651341  & 0.088888889 & 0.00187596  \\
Thriller & Closeness          & 1454175   & 0.75        & 0.733333333 & 0.011591221 \\ \rowcolor{Gray}
Thriller & Clustering         & 1374376   & 0.566666667 & 0.619047619 & 6.56E-07    \\
Thriller & Degree             & 1409367.5 & 7           & 6           & 0.000101276 \\ \rowcolor{Gray}
Thriller & Degree Centrality  & 1447459.5 & 0.666666667 & 0.648648649 & 0.006457798 \\
Thriller & Pagerank           & 1457615   & 0.115274452 & 0.110779371 & 0.015474972 \\ \rowcolor{Gray}
Thriller & Pagerank Weight    & 1432682   & 0.141588868 & 0.130622185 & 0.001550184 \\
Thriller & Total Weight       & 1494900.5 & 95          & 94          & 0.168138507 \\
 \hline
\end{tabular}
\caption{Mann–Whitney U test between men and women and by movie genre.}
\label{tab:w-test-g3}
\end{table}

\begin{table}[]
\scriptsize
\begin{tabular}{|llllll|}
\hline
Genre     & Feature            & U         & Median(M) & Median(F) & p-value     \\ \hline \rowcolor{Gray}
War      & Age Filming        & 30898.5   & 41          & 33          & 2.86E-09    \\
War      & Betweenness        & 37718.5   & 0.074074074 & 0.045687364 & 0.001715764 \\ \rowcolor{Gray}
War      & Betweenness Weight & 41149     & 0.107142857 & 0.084453782 & 0.071879433 \\
War      & Closeness          & 39490     & 0.75        & 0.695804196 & 0.015530336 \\ \rowcolor{Gray}
War      & Clustering         & 37769     & 0.571428571 & 0.660606061 & 0.001883618 \\
War      & Degree             & 39975.5   & 7           & 5           & 0.025438765 \\ \rowcolor{Gray}
War      & Degree Centrality  & 39158     & 0.666666667 & 0.591666667 & 0.010772123 \\
War      & Pagerank           & 42655     & 0.11552946  & 0.108120324 & 0.211011012 \\ \rowcolor{Gray}
War      & Pagerank Weight    & 40452     & 0.138445796 & 0.121234365 & 0.04062021  \\
War      & Total Weight       & 42494     & 75.5        & 59.5        & 0.191689163 \\ \rowcolor{Gray}
Western  & Age Filming        & 14158     & 44          & 32          & 3.93E-24    \\
Western  & Betweenness        & 22532.5   & 0.074814815 & 0.013333333 & 1.04E-08    \\ \rowcolor{Gray}
Western  & Betweenness Weight & 29978     & 0.075       & 0.066666667 & 0.053066361 \\
Western  & Closeness          & 26103.5   & 0.818181818 & 0.733333333 & 0.000119929 \\ \rowcolor{Gray}
Western  & Clustering         & 25525.5   & 0.618181818 & 0.727272727 & 3.40E-05    \\
Western  & Degree             & 29057     & 7           & 6           & 0.018342698 \\ \rowcolor{Gray}
Western  & Degree Centrality  & 25809.5   & 0.777777778 & 0.636363636 & 6.39E-05    \\
Western  & Pagerank           & 24152     & 0.12060091  & 0.096258617 & 1.24E-06    \\ \rowcolor{Gray}
Western  & Pagerank Weight    & 22135.5   & 0.155330491 & 0.099394992 & 3.53E-09    \\
Western  & Total Weight       & 29501     & 97          & 78          & 0.032522865 \\ \hline
\end{tabular}
\caption{Mann–Whitney U test between men and women and by movie genre.}
\label{tab:w-test-g4}
\end{table}
\end{document}